\documentclass{aastex62}

\usepackage{placeins}
\usepackage{morefloats}
\usepackage{graphicx}

\shorttitle{Interstellar Medium Toward Planet-Hosting Stars}
\shortauthors{Edelman et al.}

\begin{document}
\title{PROPERTIES OF THE INTERSTELLAR MEDIUM ALONG SIGHT LINES TO
NEARBY PLANET-HOSTING STARS\footnote{Based on
observations made with the NASA/ESA
Hubble Space Telescope obtained from the Data Archive at the Space
Telescope Science Institute, which is operated by the Association of
Universities for Research in Astronomy, Inc., under NASA contract NAS
AR-09525.01A. These observations are associated with programs \#12475,
12596.}}

\author[0000-0002-2764-4725]{Eric Edelman}
\affiliation{Department of Astronomy and Van Vleck Observatory, Wesleyan University, Middletown, CT 06459, USA}
\affiliation{Department of Physics and Astronomy, Embry-Riddle Aeronautical University, Prescott, Arizona 86301, USA}

\author[0000-0003-3786-3486]{Seth Redfield}
\affiliation{Department of Astronomy and Van Vleck Observatory, Wesleyan University, Middletown, CT 06459, USA}

\author[0000-0003-4446-3181]{Jeffrey L. Linsky}
\affiliation{JILA, University of Colorado and NIST, Boulder, CO 80309-0440, USA}

\author[0000-0002-4998-0893]{Brian E. Wood}
\affiliation{Naval Research Laboratory, Space Science Division, Washington, DC 20375, USA}
  
\author[0000-0001-7364-5377]{Hans M\"uller}
\affiliation{Department of Physics and Astronomy, Dartmouth College, Hanover, NH 03755, USA}

\email{eedelman@wesleyan.edu}
\email{sredfield@wesleyan.edu}

\begin{abstract}

We analyze high-resolution ultraviolet spectra of three nearby exoplanet host
stars (HD~192310, HD~9826, and HD~206860) to study interstellar
properties along their lines of sight and to search for the presence of
astrospheric absorption. Using {\em HST}/STIS spectra of the 
Lyman-$\alpha$, \ion{Mg}{2}, and
\ion{Fe}{2} lines, we identify three interstellar velocity components in
the lines of sight to each star. We can reliably assign eight of the
nine components
to partially ionized clouds found by Redfield \&
Linsky (2008) on the basis of the star's location in Galactic coordinates and
agreement of measured radial velocities with velocities predicted from
the cloud velocity vectors. None of the stars show blue-shifted
absorption indicative of an astrosphere, implying that the stars are
in regions of ionized interstellar gas. Coupling astrospheric and local interstellar medium measurements is necessary to
evaluate the host star electromagnetic and particle flux, which have 
profound impacts on the atmospheres of their orbiting planets. We
present a table of all known exoplanets located within 20~pc of the Sun
listing their interstellar properties and velocities predicted from the
local cloud velocity vectors. 

\end{abstract}


\section{Introduction}

Ultraviolet spectra of nearby late-type stars are imprinted with 
unique information 
from several distinct regions along the line of sight: (a)
emission lines and continua from the star's chromosphere and
higher temperature regions, (b) astrospheric Lyman-$\alpha$ 
absorption controlled by the stellar mass loss rate, (c) the absorption by 
interstellar gas superimposed on the stellar emission lines, and (d) heliospheric Lyman-$\alpha$ absorption controlled by the solar mass loss rate. The resulting complexity in the Lyman-$\alpha$ profile mirrors the complexity of the interaction of stars with their surrounding interstellar environments. This interface can be detected for nearby stars. Equipped with stellar wind models and measurements of the local interstellar medium (LISM), these detections can be used to estimate the relatively modest stellar winds of solar-type stars. These winds are only modest in the context of the winds of evolved stars, and can have a profound influence on the evolution of planetary atmospheres in the system.

The {\em Hubble Space Telescope} ({\em HST}), in particular with use of its Space
Telescope Imaging Spectrograph (STIS) and Cosmic Origins
Spectrograph (COS) instruments, is providing invaluable ultraviolet (UV) 
spectra critical for studying the interactions between stars, their planets, 
and the LISM. The stellar electromagnetic flux, in particular the UV flux, which can 
dominate the photochemistry and structure of planetary atmospheres 
\citep{miguel15}, has been evaluated in the context of exoplanets for 
solar-type stars of different ages \citep{Linsky2012} and M-dwarf
stars \citep{France2016}. Lyman-$\alpha$, the largest contributor of 
UV photons for late-type stars has been studied in detail by 
\citet{Wood2005b} and \citet{Youngblood2016}. Measurements of the 
intervening local interstellar medium (LISM) properties are 
necessary in order to reconstruct the intrinsic stellar Lyman-$\alpha$ line.

A spectral signature in Lyman-$\alpha$ also results from the interaction between
outflowing stellar wind protons with neutral hydrogen from the
interstellar cloud in which the star is embedded. As described in the 
now standard heliospheric model \citep[e.g.,][]{Zank}, the supersonic 
stellar wind first decelerates in a termination shock, then flows
outward and around the star in a subsonic flow ending at the astropause. 
Pristine interstellar gas is located farther from the
star beyond either the bow shock \citep{Izmodenov2003,Scherer2014} or bow wave \citep{mccomas12}. Measurements of this particle flux from stars is of prime importance when
investigating the evolution of planetary atmospheres. For example, 
\cite{lammer13} have argued that Mars lost much of its atmosphere 
from sputtering when its global magnetic field weakened once the 
planetary core cooled. The astrospheric technique for measuring
mass-loss rates from nearby late-type stars is described by 
\citet{Wood2004}.

The \textit{Voyager 1} and \textit{Voyager 2} spacecraft provided a
clear test of the heliospheric model when they crossed the solar 
termination shock 
in 2004 \citep{stone05} and 2007 \citep{stone08}, respectively, and 
have provided direct measurements of the plasma pressure and magnetic
field strength in the inner heliosheath region inside the heliopause as a function of distance from the 
Sun \citep{Decker}.
Information from the \textit{Voyagers} and 
other spacecraft, such as \textit{Ulysses}, have provided realistic 
tests of MHD and 
kinematic models of the heliosphere that focus on the interaction 
between the solar plasma and LISM neutrals
\citep[e.g.,][]{wood04review}. The \textit{Voyager} spacecraft continue to make measurements in regions that are more and more dominated by the LISM, with Voyager 1 believed to have crossed the heliopause in August of 2012 \citep{gurnett13}, while Voyager 2's crossing occurred more recently during November of 2018. \citet{zachary18} present {\it HST} observations of nearby stars along the {\it Voyager 1} and {\it Voyager 2} sight lines to compare {\it in situ} measurements of the local plasma with spectroscopic measurements of LISM absorption.
Given the ubiquity of stellar winds and surrounding interstellar 
material, these models can be generalized and applied to other stars. 
By changing a few essential parameters, heliospheric models can be
used to characterize the analogous structures around other stars, 
called astrospheres.

Within an astrosphere, the astropause is the site of charge exchange between ISM plasma and the inflowing ISM neutrals, with the neutral \ion{H}{1} from the ISM giving up 
an electron to a proton to create a population of 
very hot neutral hydrogen known as the 
``hydrogen wall''. The hydrogen
wall is centered near 150~AU for the Sun, but it can reside at different
distances from stars depending on the stellar mass-loss rate and the
relative velocity of the star compared to the interstellar flow \citep{mueller06}.
This relatively dense neutral hydrogen, which has been decelerated 
relative to the interstellar gas flow, produces broad Lyman-$\alpha$ absorption 
that is redshifted relative to the interstellar gas flow as seen from the star (e.g., the heliosphere)
but blueshifted as seen from an external perspective (e.g., astrospheres).

With high-resolution spectra, it is possible to disentangle the 
astrospheric and heliospheric absorption features from the LISM absorption in the 
Lyman-$\alpha$ line and thereby characterize the LISM along the line of
sight to a nearby star. The column density ($N($\ion{H}{1}$)$) 
of an astrospheric 
absorption feature depends on the astrosphere size and density enhancement in the hydrogen wall, which is proportional to the 
strength of the stellar wind. Therefore, a measurement of the astrospheric $N($\ion{H}{1}$)$ can be used to measure the
mass flux of a stellar wind \citep{Wood2004}, which is an 
important variable to consider when analyzing the evolution of planetary 
atmospheres and ultimately the habitability of the planets. This is the 
only observationally based technique presently available for obtaining 
mass-loss rates for solar-like and cooler stars 
\citep{Wood2014b}.  

A large stellar wind could strip an otherwise 
habitable planet of its atmosphere. It is likely that Mars lost 
its atmosphere in this fashion 4 billion 
years ago by a more active, younger Sun \citep{Wood2004,lammer13}. Moreover, in situ measurements from \textit{MAVEN} provide insights into the rates of atmospheric loss of present day Mars due, in large part, to solar influences \citep{Jakosky2018}, which, alongside evidence of Venus' atmospheric loss due to solar wind interactions in the form of its tail ray detection \citep{Grun1997} show that stellar mass-loss rates for a given star can be intimately tied to the atmospheric evolution of their planets. Obtaining 
wind-strength measurements for more solar-like stars at different 
points in their life span can provide the data needed to model the 
interactions between solar/stellar winds
and planetary atmospheres through the history of the 
solar system and other planetary systems.

Understanding the structure and kinematics of interstellar gas in the
local region of space has long been important to study general 
interstellar phenomena \citep{Frisch2011}. The local interstellar
medium (LISM) plays an important
role in this study, because local interstellar gas can be more easily
studied along the less complex short sight lines. The reduction in 
absorption profile confusion due to line blending allows the 
complexities of the LISM (e.g., abundances, dynamics, turbulence, etc.) 
to be studied in great detail. In order to measure
nearby stellar winds using the Lyman-$\alpha$ line, knowledge of the 
LISM is essential for accurately modeling the intrinsic stellar 
Lyman-$\alpha$ profile. We must also know the 
LISM flow vector (speed and direction) relative to a star, in order to
determine the three-dimensional orientation of the nose 
of the astrosphere and to estimate stellar wind
properties from the astrospheric absorption \citep{Wood2004}. Finally,
interstellar absorption decreases the detected flux in emission lines,
in particular Lyman-$\alpha$. \citet{Wood2005b} and \citet{Youngblood2016} found that absorption
by interstellar \ion{H}{1} reduces the observed flux of even nearby stars by
factors of 3--10. This is important because Lyman-$\alpha$ is the
brightest emission line in the UV spectrum of solar type stars, and the intrinsic Lyman-$\alpha$ flux is nearly as large as 
the entire UV spectra of M dwarf stars \citep{France2012}. Knowledge of
the column density and velocity structure of interstellar gas in the line of
sight to a star is needed to reconstruct stellar Lyman-$\alpha$ fluxes
\citep{Wood2005b,Linsky2013,Youngblood2016}. The intrinsic 
Lyman-$\alpha$ flux has an important influence on the photochemistry
of exoplanetary atmospheres \citep{miguel15}.

\citet{Redfield2008} have proposed a model of the LISM consisting of
15 partially ionized clouds located within 15~pc in an otherwise ionized medium.
Their model is based on high-resolution spectra of interstellar
absorption lines seen against stellar Lyman-$\alpha$,
\ion{Mg}{2}, \ion{Fe}{2}, and other UV emission lines. The spectra of 157 stars
observed with STIS and the earlier Goddard High
Resolution Spectrograph (GHRS) on {\em HST} provide radial
velocities and column densities for interstellar matter along these
sight lines. They found that the radial velocities measured in sight lines
extending over a wide angle in Galactic coordinates are consistent with
velocity vectors for these regions that define the two-dimensional 
structure of 15
clouds. While some sight lines have only a single interstellar radial 
velocity indicating a single cloud in this line of sight, many stars
located within a few parsecs show 2--4 velocity components indicating
several clouds along the line of sight. However, individual sight line 
observations cannot constrain where
along the sightline the clouds are located or whether the star at the end
of the sightline is beyond or embedded within one of the clouds. 
\citet{Redfield2015} confirm that a model of 15 bulk flow clouds is a more accurate characterization of the LISM than a small number of kinematically deformed clouds \citep{gry14}.

In this paper we analyze STIS high-resolution spectra of three nearby
stars that are known to have exoplanets. In Section~\ref{sec2} we describe the
observations and our data analysis techniques. In Section~\ref{sec3} we
evaluate the properties of LISM clouds detected in the sight lines to
these stars. In Section~\ref{sec4} we summarize the known interstellar
properties or predict these properties for all known exoplanet host
stars within 20 pc. Finally, in Section~\ref{sec5} we evaluate whether the
target stars show astrospheric absorption.

\section{Observations and Data Analysis\label{sec2}}

\subsection{Observations}
We obtained high-resolution STIS spectra of
the exoplanet-host stars HD~9826, HD~192310, and
HD~206860. HD~9826 and HD~192310 
were observed as a part of {\it HST} program 12475 ``Cool star winds and the evolution
of exoplanetary atmospheres'' (S. Redfield PI). HD~206860 
was observed as part of {\it HST} program 12596 ``In search of a young solar wind''
(B. Wood PI). Table~\ref{tab1} lists information on the three target stars. The stars
HD~35296 and HD~72905 were also observed with program 12596 and have
been analyzed by \citet{Wood2014a}, \citet{Malamut2014}, and 
\citet{Wood2014b}.

\begin{deluxetable}{lccccccc}
\tablewidth{0pt}
\tablecaption{Target Star Information. \label{tab1}}
\tablehead{Object&Distance&\textit{l}&\textit{b}&Spectral&V&$V_{R}$&Known\\
Name&(pc)&(deg)&(deg)&Type&(mag)&(km s$^{-1}$)&Planets}
\startdata
HD~192310	&	$8.91 \pm 0.02$	&$	15.6$	&$
-29.4	$&	K2~V&5.7 &$-54.2 \pm 0.90$& 2\tablenotemark{a}	\\
HD~9826	&	$13.49 \pm 0.03$	&	$132.0$	&	$-20.7$&
F9~V	&4.1 &$-28.59 \pm  0.08$&4\tablenotemark{b}\\
HD~206860	&	$17.90 \pm 0.14$	&$	69.9$&$	-28.3$	&
G0~V	&6.0&$-16.68 \pm 0.09$&1\tablenotemark{c}\\
\enddata
\tablecomments{All values are taken from the SIMBAD website unless 
otherwise noted.}
\tablenotetext{a}{\cite{Howard}, \cite{Pepe}.}
\tablenotetext{b}{\cite{Butler}, \cite{Butler2}, \cite{Curiel}.}
\tablenotetext{c}{\cite{Luhman}.}
\end{deluxetable}

\begin{figure}
\figurenum{1}
\plotone{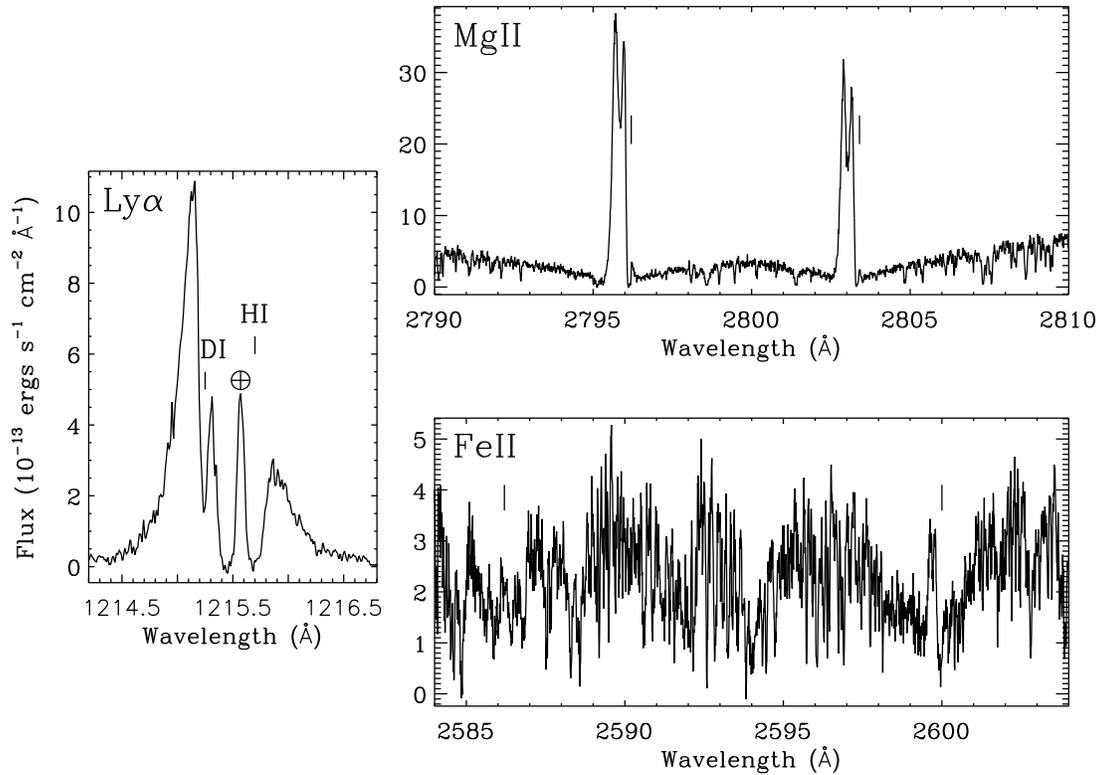}
\caption{{\em Left:} STIS Spectrum of HD~192310 showing the stellar
  Lyman-$\alpha$ emission line of \ion{H}{1} and interstellar absorption by \ion{H}{1} and
  \ion{D}{1}. Vertical lines mark the center of interstellar absorption by velocity
  component 1 and the $\oplus$ symbol marks the wavelength of the feature dominated by geocoronal emission, but which also includes a weaker contribution from the solar wind backscattered emission inseparable from the geocoronal emission in this data set. This feature is subtracted from the data prior to analysis due to its position within the portion of the line saturated due to LISM absorption. {\em Upper right:}
  The spectrum of the stellar \ion{Mg}{2} emission lines at 2796~{\AA} and
  2803~{\AA}. The vertical lines indicate absorption by interstellar
  component 1. {\em Lower right:} The STIS spectrum of the \ion{Fe}{2} 
  2586~{\AA} and 2600~{\AA} lines with 
  interstellar absorption component 1 marked.\label{fig1}}
\end{figure}

We observed the \ion{Mg}{2} 2796.3553~{\AA} and 
2803.5324~{\AA} lines and the \ion{Fe}{2} 2586.6500~{\AA} and 
2600.1729~{\AA} lines with the STIS E230H grating \citep{Hernandez2011}, 
which covered 2568--2845~{\AA} at high 
resolving power, $\lambda/\Delta\lambda=114000$. 
The \ion{Mg}{2} and \ion{Fe}{2} lines for HD~192310 
are shown in Figure~\ref{fig1}. The \ion{Mg}{2} lines have
a very high signal-to-noise  (S/N), but 
the \ion{Fe}{2} lines have a much lower S/N. 
We observed the deuterium (\ion{D}{1}) fine structure doublet at
1215.3376~{\AA} and 1215.3430~{\AA} 
and the \ion{H}{1} Lyman-$\alpha$ doublet at 1215.6682~{\AA} and 1215.6736~{\AA} 
with STIS grating E140M, which covered 
1144--1729~{\AA} at medium resolution, 
$\lambda/\Delta\lambda=45800$. 
A sample of these lines is also shown 
in Figure~\ref{fig1}. Note that the \ion{D}{1} doublet is entirely included within the 
stellar Lyman-$\alpha$ emission line.

\subsection{Fitting the \ion{Mg}{2} and \ion{Fe}{2} LISM Absorption Lines}

We began our analysis of the interstellar absorption lines of \ion{Mg}{2} and \ion{Fe}{2} using the \textit{Hubble}-specific data reduction software package known 
as \texttt{CALSTIS} \citep{Lindler1999} including the data analysis
procedures developed for the StarCAT catalog of {\it HST} stellar spectra 
\citep{ayres10}. The software
package performs dark frame subtraction, flat field calibration, flux
unit and heliocentric rest frame conversions, wavelength calibrations,
and 2-D to 1-D spectrum extraction \citep{Hernandez2011}.
Input data to our line profile fitting program includes 
the atomic parameters of each 
elemental transition taken from \cite{Morton} and our initial 
estimates of the Voigt profile parameters of each 
absorption feature. With these inputs and information regarding the 
flux and error for each pixel in the STIS spectrum,
we compute the best-fit Voigt profile 
for each absorption component 
to obtain the radial velocity of the interstellar gas 
relative to the Sun, the Doppler width of the line, and its column density.
The errors for each parameter are 
computed through Monte Carlo iterations.

\begin{figure}
\figurenum{2}
\epsscale{0.5}
\plotone{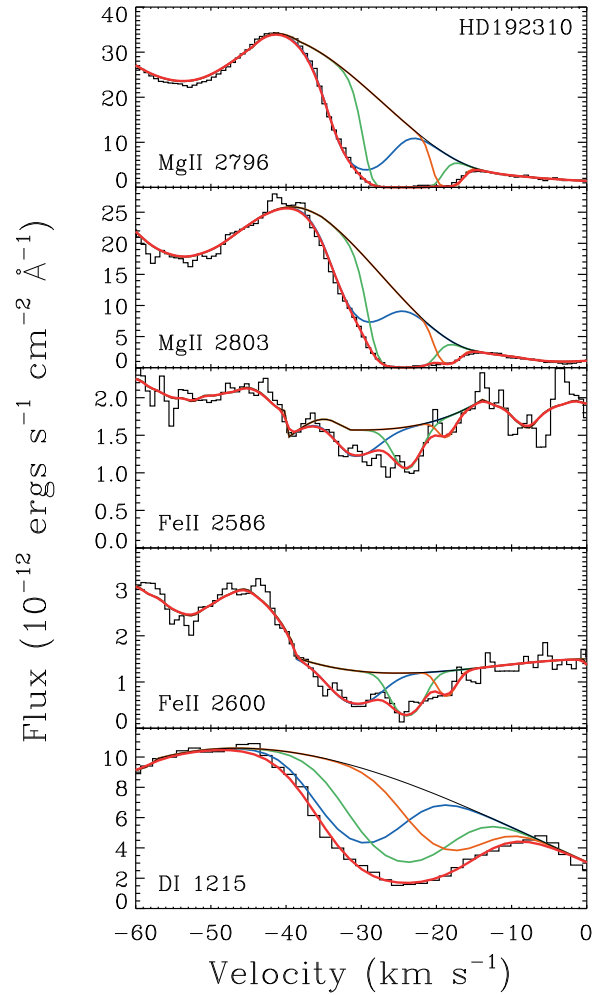}
\caption{Reconstructed stellar emission lines and interstellar \ion{Mg}{2}, \ion{Fe}{2},
and \ion{D}{1} absorption profiles for the line of sight to HD~192310. 
The blue (component 1), green (component 2), and orange (component 3) lines represent the individual interstellar absorption, while the red 
line represents the cumulative absorption by all components convolved 
with the instrumental line spread function. \ion{D}{1} is comprised of two closely spaced transitions (1215.3376~{\AA} and 1215.3430~{\AA}) and modeled as such, but combined and displayed as a single absorption for each component in the figure for clarity. 
\label{fig2}}
\end{figure}

\begin{figure}
\figurenum{3}
\epsscale{0.5}
\plotone{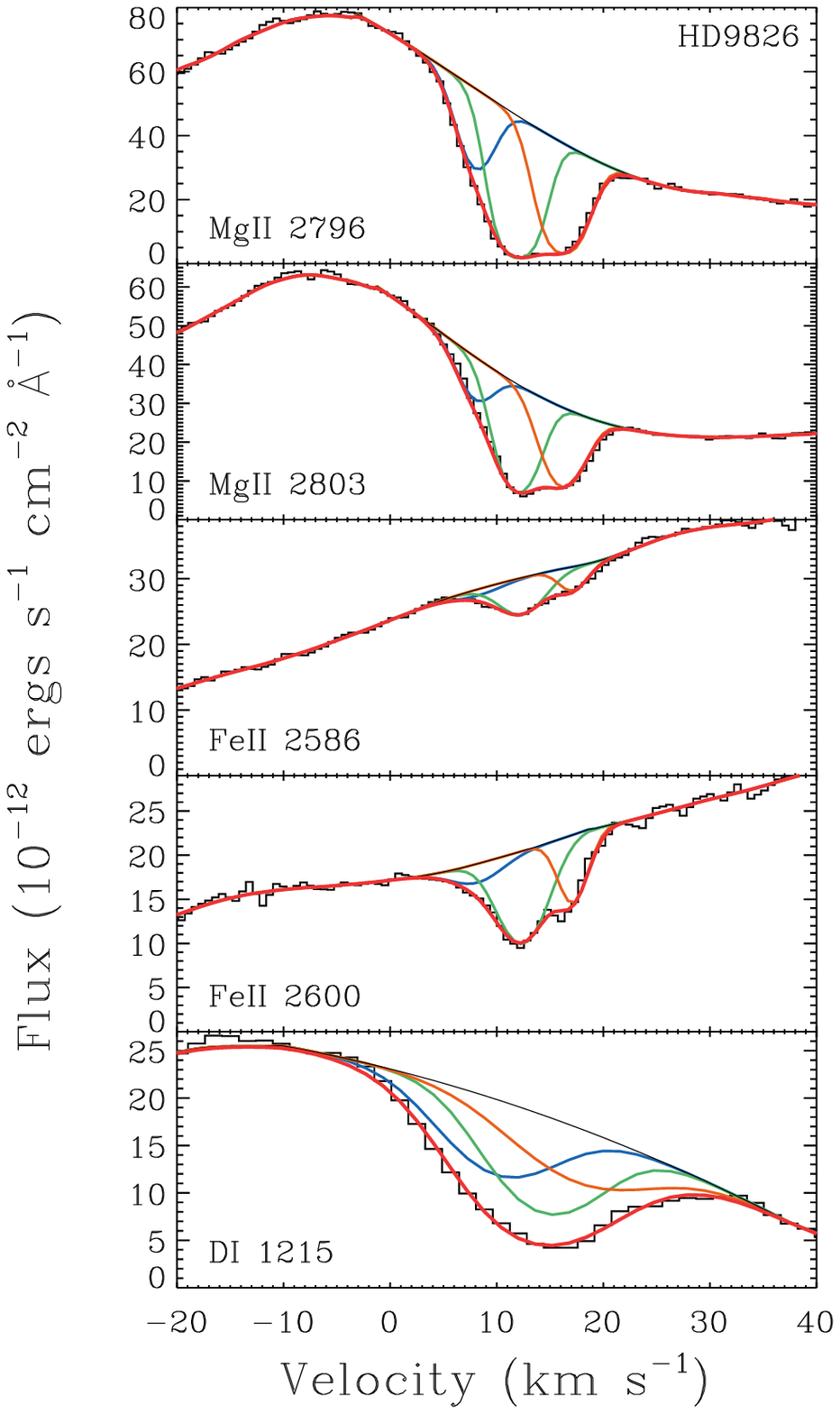}
\caption{Same as Figure~\ref{fig2} but for HD~9826.\label{fig3}}
\end{figure}

\begin{figure}
\figurenum{4}
\epsscale{0.5}
\plotone{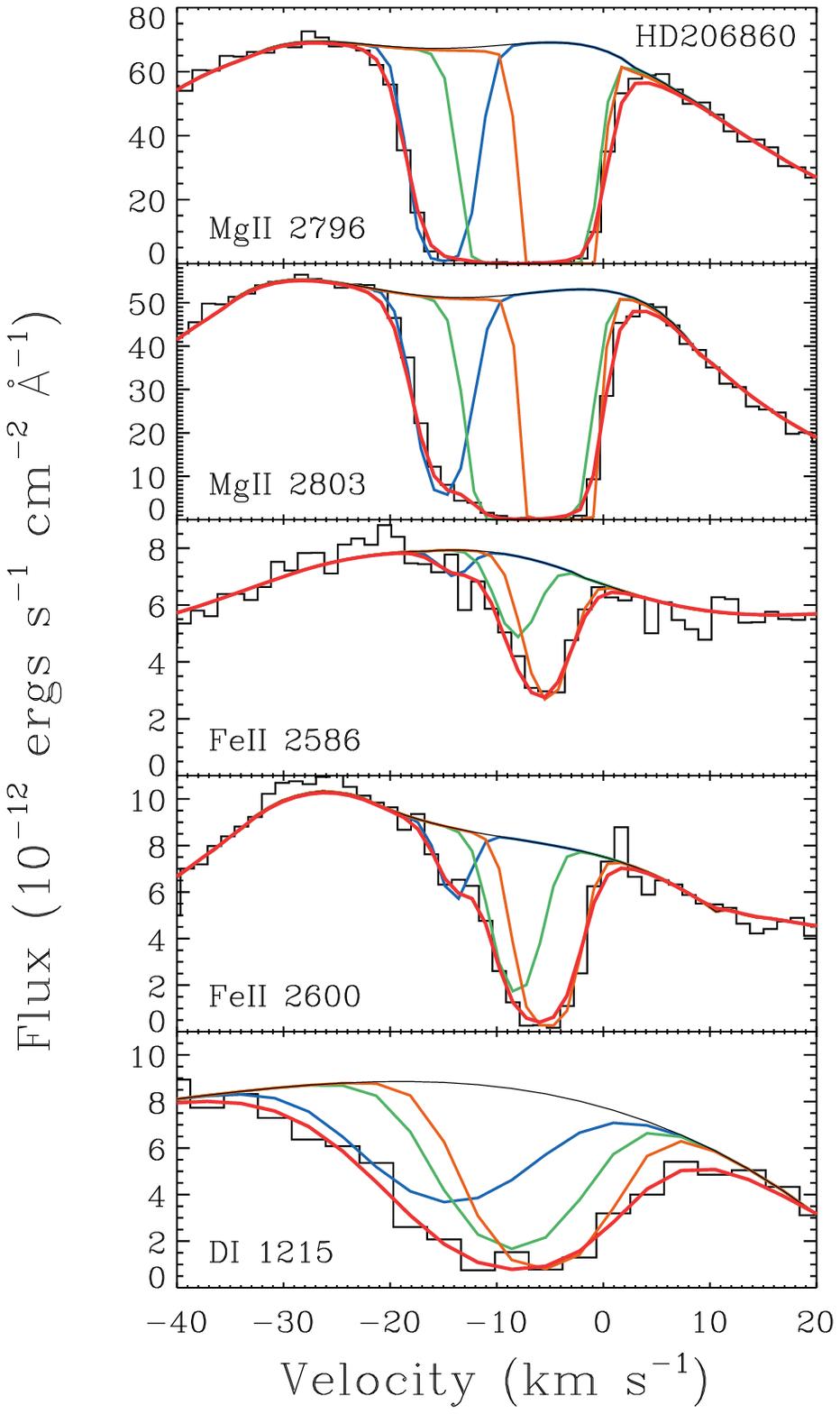}
\caption{Same as Figure~\ref{fig2} but for HD~206860.\label{fig4}}
\end{figure}

As is shown in Figures~\ref{fig2}--\ref{fig4}, the absorption profiles of the two 
\ion{Mg}{2} and two \ion{Fe}{2} lines are aligned at the same velocity 
relative to each other. This is expected because these absorption 
lines are produced by gas in the same cloud or clouds along the line 
of sight. Table~\ref{ismfit} lists the measured radial velocities, Doppler widths, 
column densities, and S/N for the three absorption components observed in the
spectra of each star. These absorption components emerged as the smallest number that could reasonably fit the feature, justified by computing an F-test for each fit with an added component. The agreement of radial velocities
measured from the different ions for each interstellar 
component mostly lie within the 1.5 km~s$^{-1}$ velocity precision of STIS.

\begin{deluxetable}{lclcccr}
\tablewidth{0pt}
\tablecaption{LISM Fit Parameters. \label{ismfit}}
\tablehead{Star &Component&Ion&Radial &Doppler&Column& S/N\\
Name&Number&&Velocity&Width&Density&\\
&&&(km s$^{-1}$)&(km s$^{-1}$)&(log N(ion))&}
\startdata
HD~192310 &	 	1	&	Fe~II	&	
$-30.69 \pm 0.66$&	$ 4.02 \pm 0.72$&$12.57 \pm 0.07$&5.4\\
    	&	 	1	&	Mg~II	&	
$-31.1 \pm 1.5$&	$3.73 \pm 0.82$&$12.20 \pm 0.09$&18.2\\
    	&	 	1	&	C~II	&	
$ -29.77 \pm 0.87$&	$6.0 \pm 1.8$&$13.46 \pm 0.16$&2.0\\
    	&	 	1	&	D~I	&	
$ -30.3 \pm 1.0$&	$7.0 \pm 1.3$&$12.85 \pm 0.20$&12.6\\
        &	 	2	&	Fe~II	&	
$ -24.04 \pm 0.51$&	$2.53 \pm 0.64$&$12.57 \pm 0.11$&5.4\\
    	&	 	2	&	Mg~II	&	
$-24.46 \pm 0.55$&	$3.76 \pm 0.60$&$13.20 \pm 0.13$&18.2\\
    	&	 	2	&	C~II	&	
$ -23.76 \pm 0.87$&	$2.4 \pm 1.0$&$15.15 \pm 0.44$&2.0\\
    	&	 	2	&	D~I	&	
$-24.9 \pm 1.0$&	$7.7 \pm 1.1$&$13.01 \pm 0.13$&12.6\\
        &	 	3	&	Fe~II	&	
$ -19.0 \pm 2.1$&	$2.5 \pm 1.9$&$12.10 \pm 0.16$&5.4\\
    	&	 	3	&	Mg~II	&	
$-18.9 \pm 1.4$&	$2.44 \pm 0.76$&$12.46 \pm 0.21$&18.2\\
    	&	 	3	&	C~II	&	
$ -18.77 \pm 0.87$&	$5.26 \pm 0.72$&$13.64 \pm 0.46$&2.0\\
    	&	 	3	&	D~I	&	
$ -19.0 \pm 1.0$&	$6.6 \pm 1.4$&$12.73 \pm 0.23$&12.6\\ \hline
	HD~9826 &	 	1	&	Fe~II	&
$10.9 \pm 1.8$&	$ 3.38 \pm 0.63$&$11.92 \pm 0.11$&21.1\\
	&	 	1	&	Mg~II	&	
$8.05 \pm 0.72$&	$2.02 \pm 0.29$&$11.57 \pm 0.13$&28.2\\
	&	 	1	&	O~I	&	
$ 9.6 \pm 1.2$&	$4.61 \pm 0.59$&$13.56 \pm 0.19$&7.5\\
	&	 	1	&	D~I	&	
$ 10.05 \pm 0.90$&	$7.2 \pm 1.3$&$12.69 \pm 0.18$&19.8\\
	&	 	2	&	Fe~II	&	
$ 12.35 \pm 0.86$&	$2.6 \pm 1.1$&$12.19 \pm 0.06$&21.1\\
	&	 	2	&	Mg~II	&	
$ 12.04 \pm 0.16$&	$2.40 \pm 0.37$&$12.49 \pm 0.05$&28.2\\
	&	 	2	&	O~I	&	
$ 13.6 \pm 1.2$&	$2.4 \pm 1.1$&$13.57 \pm 0.60$&7.5\\
	&	 	2	&	D~I	&	
$14.33 \pm 0.90$&	$6.7 \pm 1.5$&$12.88 \pm 0.16$&19.8\\
	&	 	3	&	Fe~II	&	
$ 16.67 \pm 0.79$&	$2.04 \pm 0.71$&$12.02 \pm 0.09$&21.1\\
	&	 	3	&	Mg~II	&	
$16.30 \pm 0.21$&	$2.47 \pm 0.17$&$12.38 \pm 0.05$&28.2\\
	&	 	3	&	O~I	&	
$ 17.6 \pm 1.2$&	$2.1 \pm 1.2$&$14.5 \pm 1.1$&7.5\\
	&	 	3	&	D~I	&	
$ 18.24 \pm 0.90$&	$8.8 \pm 1.9$&$12.70 \pm 0.16$&19.8\\ \hline
HD~206860 &	 	1	&	Fe~II	&	
$-13.80 \pm 0.98$&	$ 2.35 \pm 0.71$&$11.99 \pm 0.09$&11.4\\
&	 	1	&	Mg~II	&	
$-14.77 \pm 0.41$&	$2.59 \pm 0.17$&$12.67 \pm 0.04$&30.3\\
&	 	1	&	D~I	&	
$ -14.84 \pm 0.67$&	$9.7 \pm 1.6$&$13.05 \pm 0.13$&10.4\\
&	 	2	&	Fe~II	&	
$ -7.14 \pm 0.91$&	$2.47 \pm 0.39$&$12.69 \pm 0.12$&11.4\\
&	 	2	&	Mg~II	&	
$ -7.43 \pm 0.69$&	$3.47 \pm 0.27$&$13.22 \pm 0.28$&30.3\\
&	 	2	&	D~I	&	
$-9.00 \pm 0.67$&	$7.1 \pm 1.9$&$13.19 \pm 0.38$&10.4\\
&	 	3	&	Fe~II	&	
$ -5.00 \pm 0.65$&	$2.11 \pm 0.51$&$12.89 \pm 0.08$&11.4\\
&	 	3	&	Mg~II	&	
$-4.87 \pm 0.82$&	$2.07 \pm 0.30$&$14.16 \pm 0.20$&30.3\\
&	 	3	&	D~I	&	
$ -6.29 \pm 0.67$&	$6.5 \pm 1.5$&$13.31 \pm 0.23$&10.4
\enddata
\tablecomments{S/N based on an average of 10 data points over the 
emission line near absorption.}
\end{deluxetable}

\subsection{Fitting the \ion{H}{1} and \ion{D}{1} Lyman-$\alpha$ Absorption Lines}

\begin{figure}
\figurenum{5}
\epsscale{0.9}
\plotone{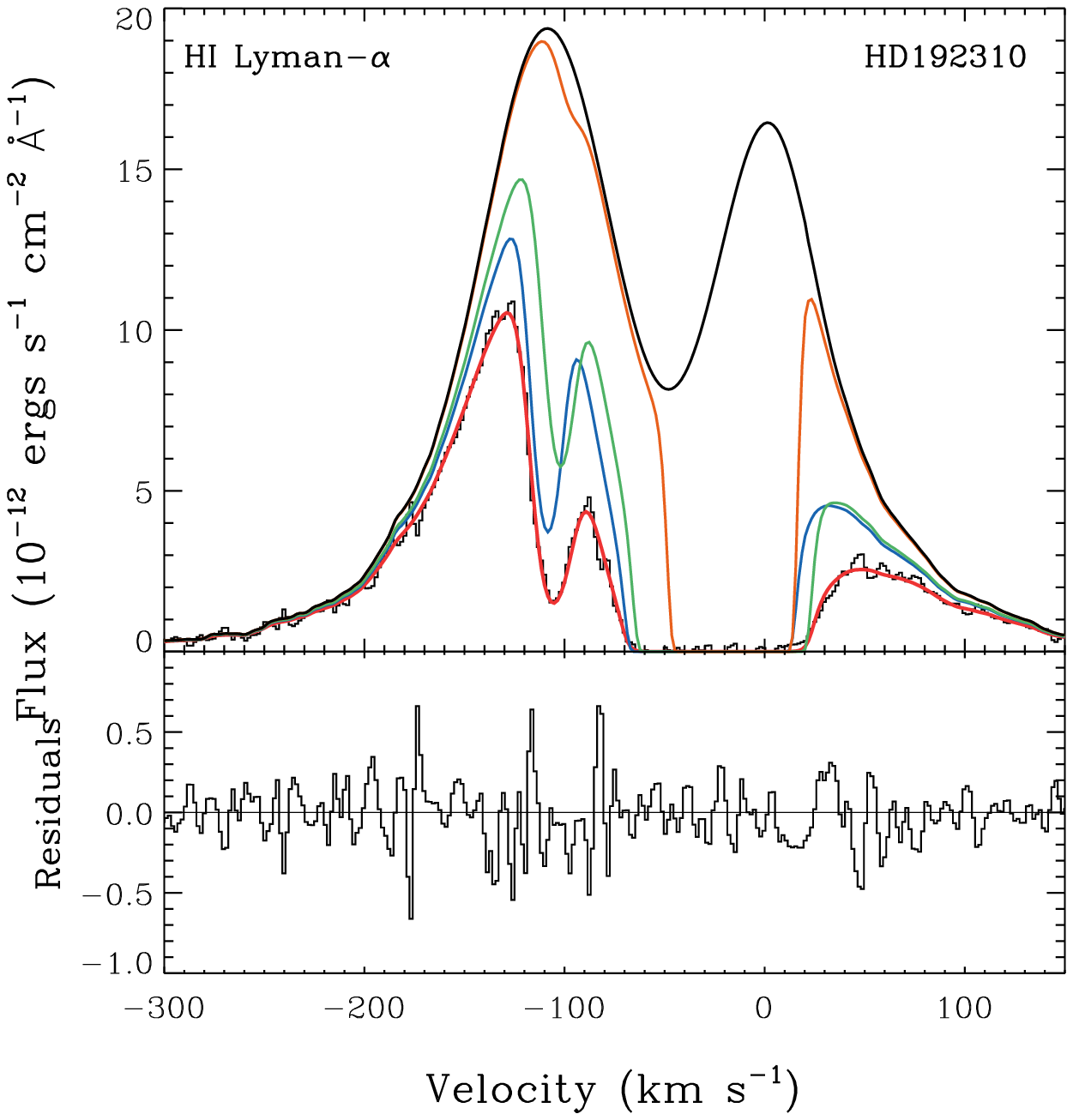}
\label{fig5}
\caption{Reconstructed stellar \ion{H}{1} Lyman-$\alpha$ emission line (black
line) for HD~192310 and interstellar absorption by component 1 (blue
line), component 2 (green line), and component 3 (orange
line). For each of the three interstellar components there are four 
absorption lines as the Lyman-$\alpha$ lines of both \ion{H}{1} and \ion{D}{1} are close 
doublets. However, for clarity, these have all been combined and displayed as a single absorption feature for each component. The center of \ion{D}{1} absorption is at --81.6 km~s$^{-1}$ relative to \ion{H}{1}.
The thick red line is the convolution of the intrinsic Lyman-$\alpha$ emission line with the sum of the interstellar absorption components. It is an excellent fit to the observed 
profile (thin black histogram). The difference in flux versus 
the total predicted absorption is plotted in velocity space beneath
the fit.}
\end{figure}

\begin{figure}
\figurenum{6}
\epsscale{0.9}
\plotone{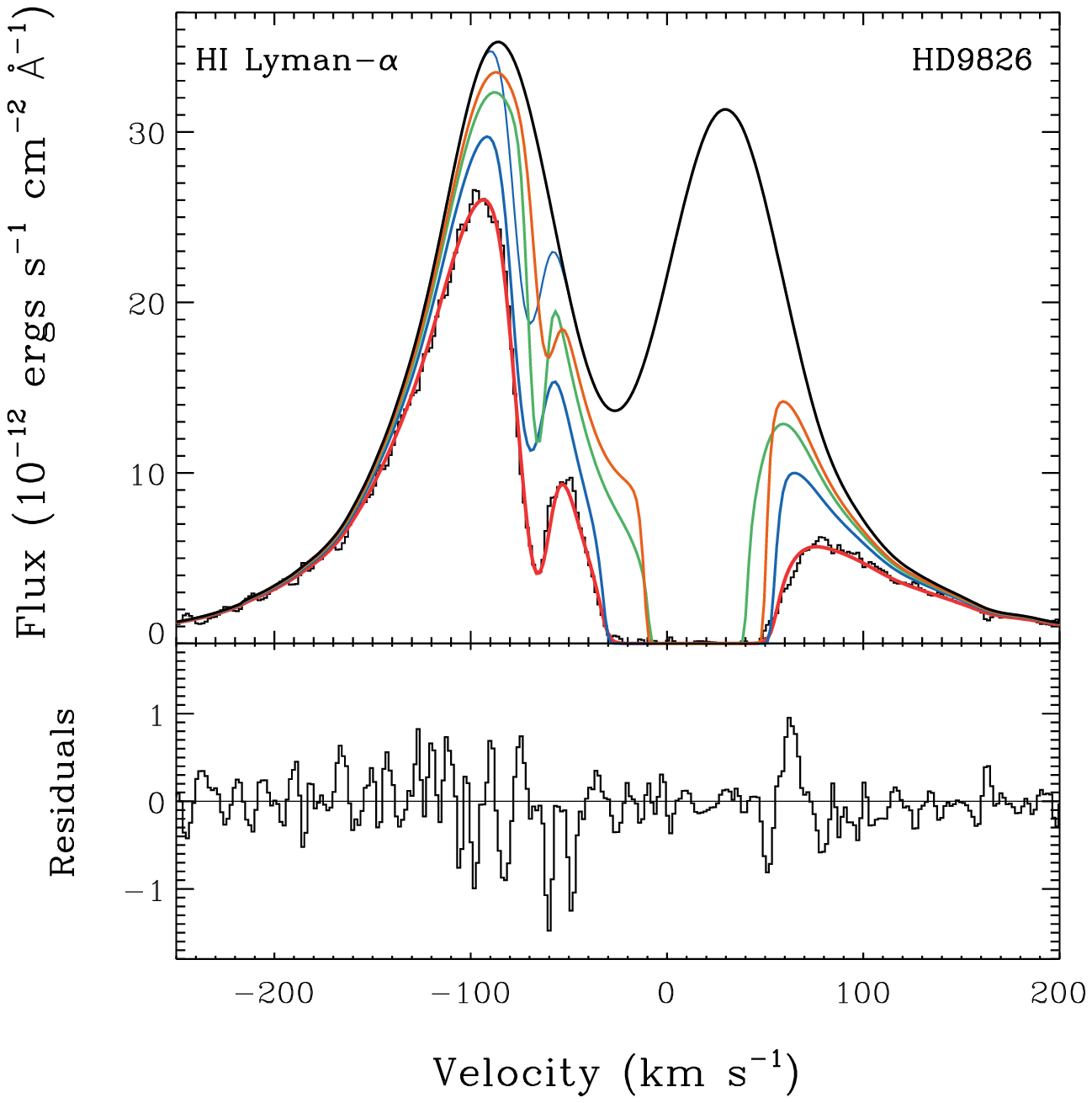}
\label{fig6}
\caption{Same as Figure~\ref{fig5} but for HD~9826.}
\end{figure}

\begin{figure}
\figurenum{7}
\epsscale{0.9}
\plotone{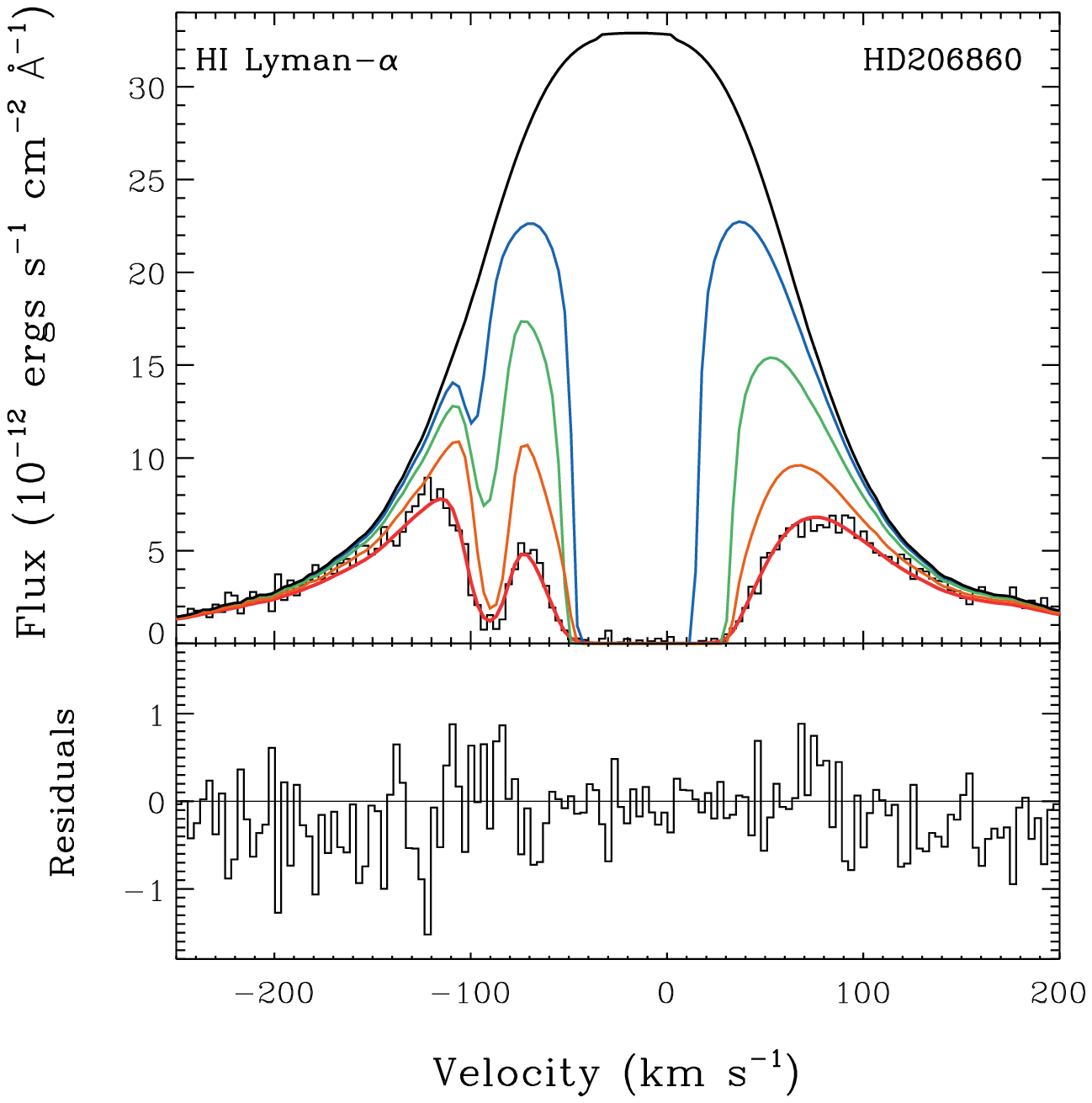}
\label{fig7}
\caption{Same as Figure~\ref{fig5} but for HD~206860.}
\end{figure}

We next fitted the interstellar Lyman-$\alpha$ absorption lines of atomic 
hydrogen and deuterium. This procedure differs somewhat from that used
to fit the \ion{Mg}{2} and \ion{Fe}{2} lines, because interstellar absorption
removes the core of the intrinsic stellar Lyman-$\alpha$ emission
line, requiring a reconstruction of the intrinsic emission profile. 
Following the fitting procedure described by
\citet{Wood2005,Wood2014a}, we first assume that the absorption velocities of
\ion{H}{1} and \ion{D}{1} are the same and that thermal broadening dominates the width of the \ion{H}{1} and \ion{D}{1}
lines, and therefore differ only due to
the factor of two difference in mass. Our fit to the narrow \ion{D}{1}
line measures the deuterium column density $N($\ion{D}{1}$)$, which
provides an estimate of the hydrogen column density $N($\ion{H}{1}$)$ given the
number density ratio D/H = $1.56\times10^{-5}$ in the LISM
\citep{linsky06}. This value for $N($\ion{H}{1}$)$ provides a good estimate 
of the shape of the
intrinsic Lyman-$\alpha$ emission line wings. We also assume that the shape of the
Lyman-$\alpha$ line core is similar to that of the \ion{Mg}{2} lines 
including a self-reversal,
because both the \ion{Mg}{2} and Lyman-$\alpha$ lines are optically thick and
are formed in the stellar chromosphere.
We then find the best fit to the observed Lyman-$\alpha$ profile by allowing small 
changes in the intrinsic 
emission line shape and relaxing the assumption that the \ion{H}{1} and \ion{D}{1}
velocities must be the same.
The results of these fits are shown in Figures~\ref{fig5}--\ref{fig7} for HD~192310, HD~9826, and 
HD~206860, and the resulting radial velocities, Doppler
widths, and column densities are listed in Table~\ref{laismfit}.

\begin{deluxetable}{lccccr}
\tablewidth{0pt}
\tablecaption{Lyman-$\alpha$ ISM Fit Parameters. \label{laismfit}}
\tablehead{Object &Component&Radial &Doppler&Column& S/N\\
\hspace{0cm} Name&Number&Velocity&Width&Density&\\
\hspace{1cm}&&(km s$^{-1}$)&(km s$^{-1}$)&(log N(H~I))}
\startdata
HD~192310 &	 	1	&	$-27.47 \pm 0.45 $&
$ 12.69 \pm 0.66$&$17.96 \pm 0.03$&8.0\\
&	 	2	&	$-21.48 \pm 0.45$&	
$13.45 \pm 0.75$&$17.90 \pm 0.03$&8.0\\
&	 	3	&	$-16.47 \pm 0.45$&	
$11.5 \pm 1.4$&$16.17 \pm 0.50$&8.0\\ \hline
		HD~9826 &	 	1	&	
$10.83 \pm 1.13$&	$ 13.29 \pm 0.30$&$17.78 \pm 0.20$&12.2\\
    	&	 	2	&	$15.11 \pm 1.13$&
$7.53 \pm 2.19$&$17.55 \pm 0.17$&12.2\\
    	&	 	3	&	$19.02 \pm 1.13$&
$9.80 \pm 1.74$&$17.34 \pm 0.28$&12.2\\ \hline
HD~206860 &	 	1	&	$-16.6 \pm 1.2$&
$ 10.0 \pm 1.5$&$17.35 \pm 0.26$&9.7\\
&	 	2	&	$-10.8 \pm 1.2$&	
$13.0 \pm 1.1$&$17.79 \pm 0.20$&9.7\\
&	 	3	&	$-8.1 \pm 1.2$&	$12.0 \pm1.5$&
$18.14 \pm 0.17$&9.7\\
\enddata
\end{deluxetable}

Since the Lyman-$\alpha$ lines of both \ion{H}{1} and \ion{D}{1} are fine-structure 
doublets separated by 0.0054~{\AA}, corresponding to 1.33 km~s$^{-1}$, 
absorption by each LISM cloud appears as two closely spaced 
absorption components for each isotope. For all three stars, we find from the \ion{Mg}{2} and \ion{Fe}{2} absorption spectra that spectra for
the sight lines to each star show interstellar absorption at three different velocities. We must, therefore, 
fit each \ion{H}{1} and \ion{D}{1} Lyman-$\alpha$ ISM velocity component with three free parameters translating to
12 absorption components. Figures~\ref{fig5}--\ref{fig7} show the interstellar
absorption profiles for each line of sight, where we combined the closely spaced fine structure lines for clarity. Also shown is the combined line profile convolved with the instrumental line spread function,
and the observed spectrum.

We find satisfactory agreement in the interstellar radial velocity between the line
fits of \ion{D}{1}/\ion{H}{1} and \ion{Mg}{2} and \ion{Fe}{2}. We
conclude, therefore, that interstellar absorption alone without any additional
absorption components (e.g., astrospheric or heliospheric absorption) can explain these
observations. 

\section{Properties of the LISM along the Observed Sight Lines\label{sec3}}

The LISM is a complex environment filled with a variety of 
partially ionized interstellar gas clouds that produce
absorption lines superimposed on the stellar emission line spectra. Ultraviolet 
resonance lines of \ion{H}{1}, \ion{D}{1}, O~I, C~II, \ion{Fe}{2}, \ion{Mg}{2}, and other atoms
and ions are typically observed in high-resolution spectra of even the
closest stars \citep[e.g.,][]{Redfield2002,Redfield2004A}. Figures~\ref{fig2}--\ref{fig4}
show our fits to the major ISM features in the data sets for 
HD~192310, HD~9826, and HD~206860, respectively. We focus our attention
on the high signal-to-noise absorption features of \ion{Mg}{2}, \ion{Fe}{2}, \ion{D}{1}, and 
of course, \ion{H}{1} Lyman-$\alpha$.

We compare our measurements with the 
\cite{Redfield2008} dynamical model of the LISM. This model uses 
information gathered from GHRS and STIS observations of 157 sight lines to compute
velocity vectors of 15 LISM clouds located within 15~pc of the Sun. For specified 
Galactic coordinates, the model predicts which clouds 
intersect or lie within $20^{\circ}$ of the sightline to the star, and the
 radial velocities for these clouds 
\citep{Redfield2008}\footnote{The LISM Kinematic Calculator: 
\url{http://lism.wesleyan.edu/LISMdynamics.html}}. In Table~\ref{propertiestable} we list the interstellar clouds that best match the three
velocity components for each line of sight. We select the most likely
clouds on the basis that (1) the line of sight is inside, at the edge, or
close to the edge of a known cloud, and (2) the radial velocity predicted 
from the cloud's velocity vector lies within 
2.0~km~s$^{-1}$ of the weighted mean velocity of the observed narrow
interstellar lines (\ion{Mg}{2}, \ion{Fe}{2}, \ion{D}{1}, and either C~II or O~I). 
The mean velocities weighted by the inverse square of the measured
errors of the individual velocities from Table~\ref{ismfit} for each ion are listed in Table~\ref{propertiestable}.
Given these criteria, we can
reliably match 8 of the 9 velocity components with one or more clouds.
In one case (component 1 for HD~206860) there are two plausible cloud
identifications, although the line of sight passes outside of the
presently known boundaries of both the Vel and Mic clouds.
There is no known cloud that matches the
line of sight to HD~9826 with a velocity close to that of component 1.

\begin{figure}
\figurenum{8}
\includegraphics[scale=0.7,angle=90]
{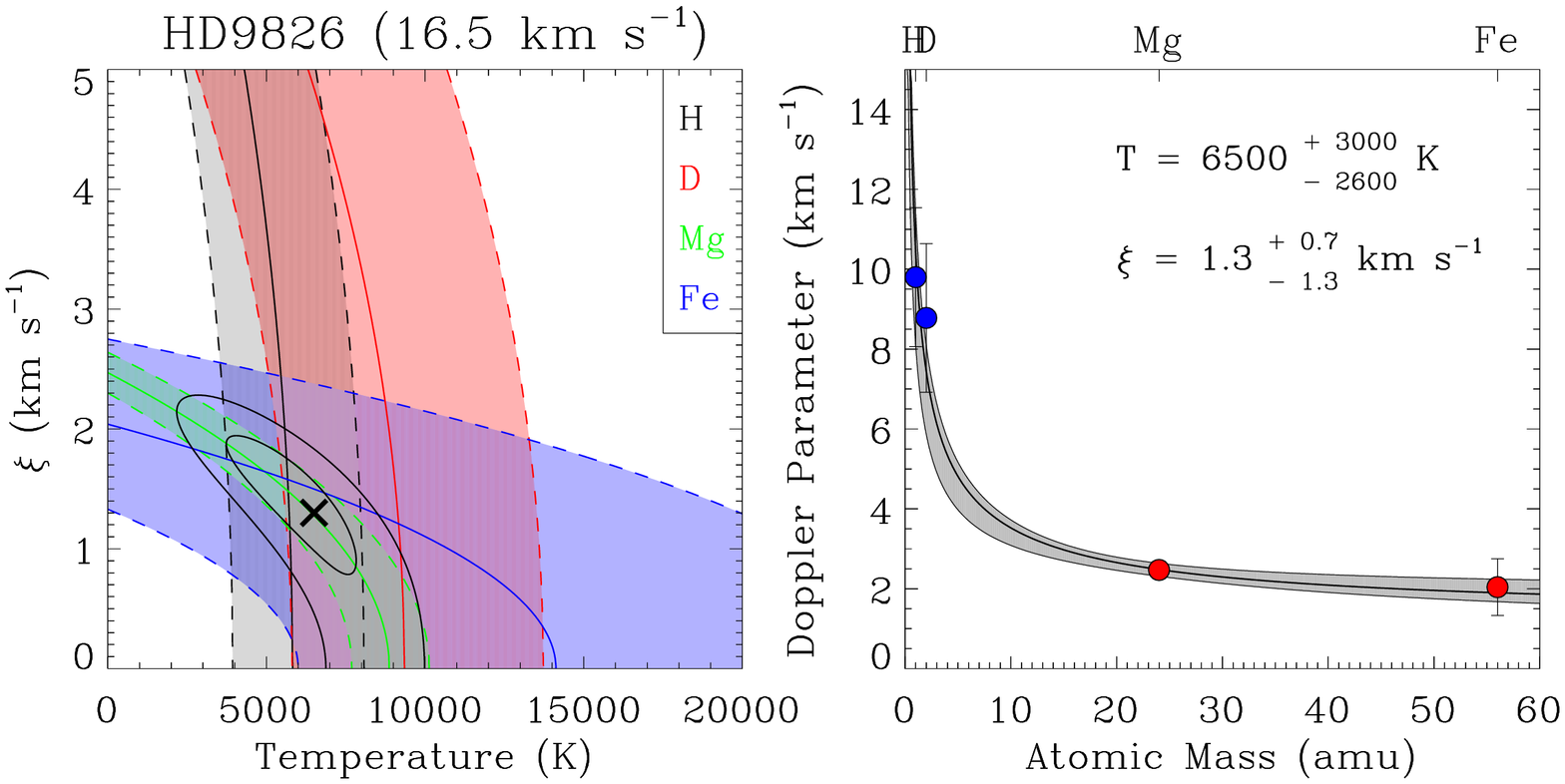}
\label{fig8}
\caption{Temperature and
turbulent velocity fits for component 3 ($\approx$16.5 km~s$^{-1}$ for ions listed in Table~\ref{ismfit} and $\approx$19.0 km~s$^{-1}$ for Lyman-$\alpha$) in the line of sight to HD~9826.
{\em Left:} Solid lines plot the best fit temperature-turbulence
relation for \ion{H}{1} (black), \ion{D}{1} (red), \ion{Mg}{2} (green), and \ion{Fe}{2} (red) and
dashed lines show the 1$\sigma$ uncertainties in these fits. The
$\times$ symbol marks the best fit considering all four elements and the
curved lines around the $\times$ symbol mark the 1$\sigma$ and 2$\sigma$
uncertainties. {\em Right:} Plot of the Doppler width (thermal and
turbulent combined) vs. atomic mass for the four elements. The dashed
lines above and below the solid line mark the 1$\sigma$ uncertainty 
about the mean fit (solid line).
The derived uncertainties in the temperature and turbulent velocity
are 1$\sigma$ values.}

\end{figure}

As the measured Doppler width parameter $b$ is equal to the sum of the thermal and non-thermal broadening processes, $b$ can be used to provide reasonable estimations of
temperature ($T$) 
and turbulence ($\xi$) of the absorbing gas. This is accomplished with the use of the following equation \citep{Draine} that adds together the two broadening factors
\begin{equation}
b^{2} = \frac{2kT}{m} + \xi^2,
\end{equation} 
and by taking into account the different
ion masses. Figure~\ref{fig8} is an example of such a measurement for component 3 detected toward HD~9826. See \citet{Redfield2004B} for a more detailed discussion of this technique. In all cases, the agreement of cloud temperature, listed in \citet{Redfield2008}, with
the measured temperature lies within the admittedly large errors, as displayed in Table~\ref{propertiestable}.

We measure the depletions (in log units) for iron and magnesium
by comparison with the solar abundances obtained by \cite{Asplund},
assuming that they are completely in their first ionization state, which is a reasonable assumption \citep{slavin08}. 
The values of these properties for the three 
lines of sight considered in this paper can be found in Table~\ref{propertiestable}, and in most cases are reasonably consistent with the associated cloud averages from \citet{Redfield2008}. 
These physical properties of the surrounding interstellar medium (e.g., velocity, temperature, and ionization) become extremely important when creating 
model astrospheres for stars.

\begin{deluxetable}{lcccccc}
\tabletypesize{\small}
\tablecaption{Physical Properties of the ISM Towards Planet-Hosting Stars\tablenotemark{a}\label{propertiestable}}
\tablewidth{0pt}

\tablecolumns{7}
\tablehead{
\colhead{Star} & \colhead{Component} & \colhead{$v$} &
\colhead{$T$} & \colhead{$\xi$}
& \colhead{$D$(Mg)} & \colhead{$D$(Fe)}\\
\colhead{Name} & \colhead{or Cloud} &
\colhead{(km~s$^{-1}$)} &  \colhead{(K)} & 
\colhead{(km s$^{-1}$)} & \colhead{} & \colhead{}}
\startdata
HD~192310 & 1 & $-30.41 \pm 0.57$ & $8600^{+2000}_{-1800}$ & $3.3^{+1.2}_{-1.3}$ &
   $-1.29\pm0.13$ & $-0.84\pm0.09$\\
   & Vel edge & $-31.70\pm1.52$ & (10600) & (3.5) & & (--0.03)\\
   & 2 & $-24.24 \pm 0.50$ & $9900^{+2200}_{-2100}$ & $2.2^{+1.1}_{-2.0}$ &
   $-0.23\pm0.16$ & $-0.78\pm0.12$\\
   & Mic inside & $-23.74\pm1.02$ & $9900\pm2000$ & $3.1\pm1.0$ & 
   $-0.03\pm0.40$ & $-0.92\pm0.43$\\
   & 3 & $-18.88 \pm 0.57$ & $6900^{+2600}_{-2300}$ & $1.3^{+1.6}_{-1.3}$ & $-0.76\pm0.55$
   & $-0.48\pm0.53$\\
   & LIC outside & $-17.32\pm1.15$ & $7500\pm1300$ & $1.62\pm0.75$ & 
   $-0.97\pm0.23$ & $-1.12\pm0.10$\\ \hline
HD~9826 & 1 & $9.1 \pm 1.2$ & $10400^{+2000}_{-1900}$ & $0.00^{+2.4}_{-0.0}$ &
   $-1.74\pm0.25$ & $-1.31\pm0.23$\\
   & unknown & & & & & \\
   & 2 & $12.1 \pm 1.1$ & $4000^{+2800}_{-2200}$ & $1.8^{+0.8}_{-1.2}$ & $-0.59\pm0.20$ &
   $-0.81\pm0.19$\\ 
   & Hyades inside & $12.06\pm1.09$ & $6200\pm3800$ & $2.7\pm1.2$ & 
   $-1.06\pm0.47$ & $-0.32\pm0.62$\\
   & 3 & $16.45 \pm 0.88$ & $6500^{+3000}_{-2600}$ & $1.3^{+0.7}_{-1.3}$ & $-0.49\pm0.30$ &
   $-0.77\pm0.30$\\
   & LIC inside & $14.44\pm1.23$ & $7500\pm1300$ & $1.62\pm0.75$ & 
   $-0.97\pm0.23$ & $-1.12\pm0.10$\\  \hline
HD~206860 & 1 & $-14.68 \pm 0.58$ & $7100^{+2800}_{-2400}$ & $1.4^{+0.6}_{-1.4}$ &
   $-0.21\pm0.28$ & $-0.81\pm0.28$\\
   & Vel outside & $-15.36\pm0.99$ & (10600) & (3.5) & & (--0.03)\\
   & Mic outside & $-16.33\pm1.34$ & $9900\pm2000$ & $3.1\pm1.0$ & 
   $-0.03\pm0.40$ & $-0.92\pm0.43$\\
   & 2  & $-8.0 \pm 1.0$ & $9600^{+2500}_{-2300}$ & $2.11^{+0.54}_{-0.68}$ &
   $-0.10\pm0.36$ & $-0.55\pm0.24$\\
   & Eri inside & $-8.65\pm0.99$ & $5300\pm4000$ & $3.6\pm1$ & 
   $-0.15\pm0.30$ & $-0.39\pm0.19$\\
   & 3 & $-5.44 \pm 0.79$ & $6800^{+2700}_{-2500}$ & $0.80^{+0.98}_{-0.80}$ &
   $-0.49\pm0.28$ & $-0.70\pm0.19$\\
   & LIC edge & $-6.70\pm1.35$ & $7500\pm1300$ & $1.62\pm0.75$ & 
   $-0.97\pm0.23$ & $-1.12\pm0.10$\\
\enddata
\tablenotetext{a}{Sight line velocities are the weighted means of the heavy ions measured in Table~\ref{ismfit}. The errors are the larger of either the error on the weighted mean or the standard deviation of the individual values. The corresponding clouds are indicated by name and whether the sightline is inside, at the edge (i.e., $<$20$^\circ$), or well outside (i.e., $>$20$^\circ$) of the cloud boundaries identified by Redfield \& Linsky (2008). For HD~9826, component 1 corresponds to no known cloud. For HD~206860, component 1 is more likely the Vel Cloud but could be the Mic Cloud. Cloud parameters in parenthesis are based on only one line of sight.}
\end{deluxetable}

\section{Properties of the LISM Clouds in the Sight Lines to Known
  Planet-Hosting Stars Within 20 pc of the Sun\label{sec4}}

    The kinematic and physical properties of the LISM clouds are critical input parameters for analyzing the profiles of stellar emission lines and for measuring the mass-loss rates of host star winds that can erode exoplanet atmospheres. While spectra of Lyman-$\alpha$ and other emission lines obtained during transits provide valuable transmission spectra \citep[e.g.,][]{vidalmadjar03, fossati10, ehrenreich15, loyd17} that can be used to measure the extent and composition of exoplanet atmospheres, interstellar absorption in the line cores severely complicates the analysis. It is essential, therefore, to know and remove the interstellar absorption components in these lines. Spectroscopic follow-up using ground-based telescopes, {\em HST}, and the {\em James Webb Space Telescope} of bright nearby stars with transiting exoplanets discovered by ground and space-based facilities, including the recently launched {\em Transiting Exoplanet Survey Satellite} ({\it TESS}) mission will result in powerful studies of exoplanetary atmospheres. To facilitate such studies, we include in Table~\ref{planettable} a summary of the measured and predicted LISM properties for the lines of sight to all presently known exoplanet host stars located within 20 pc of the Sun. This currently totals 96 systems, eight of which contain transiting exoplanets. We compiled the list from the Confirmed Planets Table at the NASA Exoplanet Archive\footnote{\url{https://exoplanetarchive.ipac.caltech.edu}} supplemented by Gaia DR2\footnote{\url{https://gea.esac.esa.int/archive/}} \citep{Gaia2016,Gaia2018} and SIMBAD\footnote{\url{http://simbad.u-strasbg.fr}}. We list the
measured radial velocities and $log($\ion{H}{1}$)$ column densities for each identified
interstellar velocity component, together with references to the
original papers.

There are no measured interstellar properties for
most of the stars in the list and therefore we rely on the morphological \citep{Redfield2000} and kinematic \citep{Redfield2008} models of the LISM. The clouds 
traversed by each line of sight to these stars are those for which the 
Galactic coordinates are inside of or very
close to the cloud boundaries shown in
\cite{Redfield2008}. We compute the predicted cloud velocities from the cloud
velocity vectors\footnote{\url{http://lism.wesleyan.edu/LISMdynamics.html}}, and
for lines of sight through the Local Interstellar Cloud (LIC) we compute the \ion{H}{1} column densities
from the three-dimensional LIC model \citep{Redfield2000}\footnote{\url{http://lism.wesleyan.edu/ColoradoLIC.html}}. The first component
is likely the strongest absorber. Note that the predicted velocities 
for several clouds along the same line of sight are similar in many
cases. Therefore, low-resolution spectra for these lines of sight will 
likely show a blended profile of interstellar absorption.  

Because strong emission lines, e.g. \ion{H}{1} Lyman-$\alpha$ (1216~\AA), 
O~I (1302~\AA), C~II (1334~\AA), and \ion{Mg}{2} (2796~\AA\ and 2802~\AA),
are transitions to the ground state, absorption by interstellar atoms and ions, which are always in their ground states, near line center alters the observed profile in ways that can almost completely obscure or significantly alter the signal produced by the exoplanet. The local region of
space ($<$100\,pc) occupied by the LISM, is also the same volume of the galaxy as the most attractive exoplanets for transmission spectroscopy follow-up. For this reason, it is important to have as complete as possible knowledge on the properties of the LISM.

\section{Search for Spectroscopic Evidence of Astrospheres and the Heliosphere\label{sec5}}

We selected the stars in this study
because they are host stars of confirmed exoplanets. 
The detection of astrospheric absorption from any of the stars would be a great opportunity to infer 
the mass loss rate from the host star of an exoplanet.

The signature of an astrosphere is additional absorption by 
\ion{H}{1} gas in the star's ``hydrogen wall'' produced
by charge-exchange interactions 
between the ISM plasma slowed by the outflowing stellar wind and the neutral hydrogen in the interstellar cloud that embeds the star. These interactions lead to Lyman-$\alpha$ absorption that is red-shifted relative to the inflowing gas as seen from the
star but blue-shifted as seen by an external observer. We have, therefore,
inspected the fitted Lyman-$\alpha$ absorption lines to look for
additional absorption on the short wavelength side of the interstellar 
\ion{H}{1} absorption. Figures~\ref{fig5}--\ref{fig7} show
details of the Lyman-$\alpha$
lines of all three stars, comparing our fit to the observed
Lyman-$\alpha$ lines that includes interstellar \ion{H}{1} and \ion{D}{1} absorption by
all 3 velocity components along their sight lines. These figures show
no evidence for additional absorption at shorter wavelengths than the
interstellar \ion{H}{1} absorption and thus no evidence for astrospheric absorption.

\begin{figure}
\figurenum{9}
\plottwo{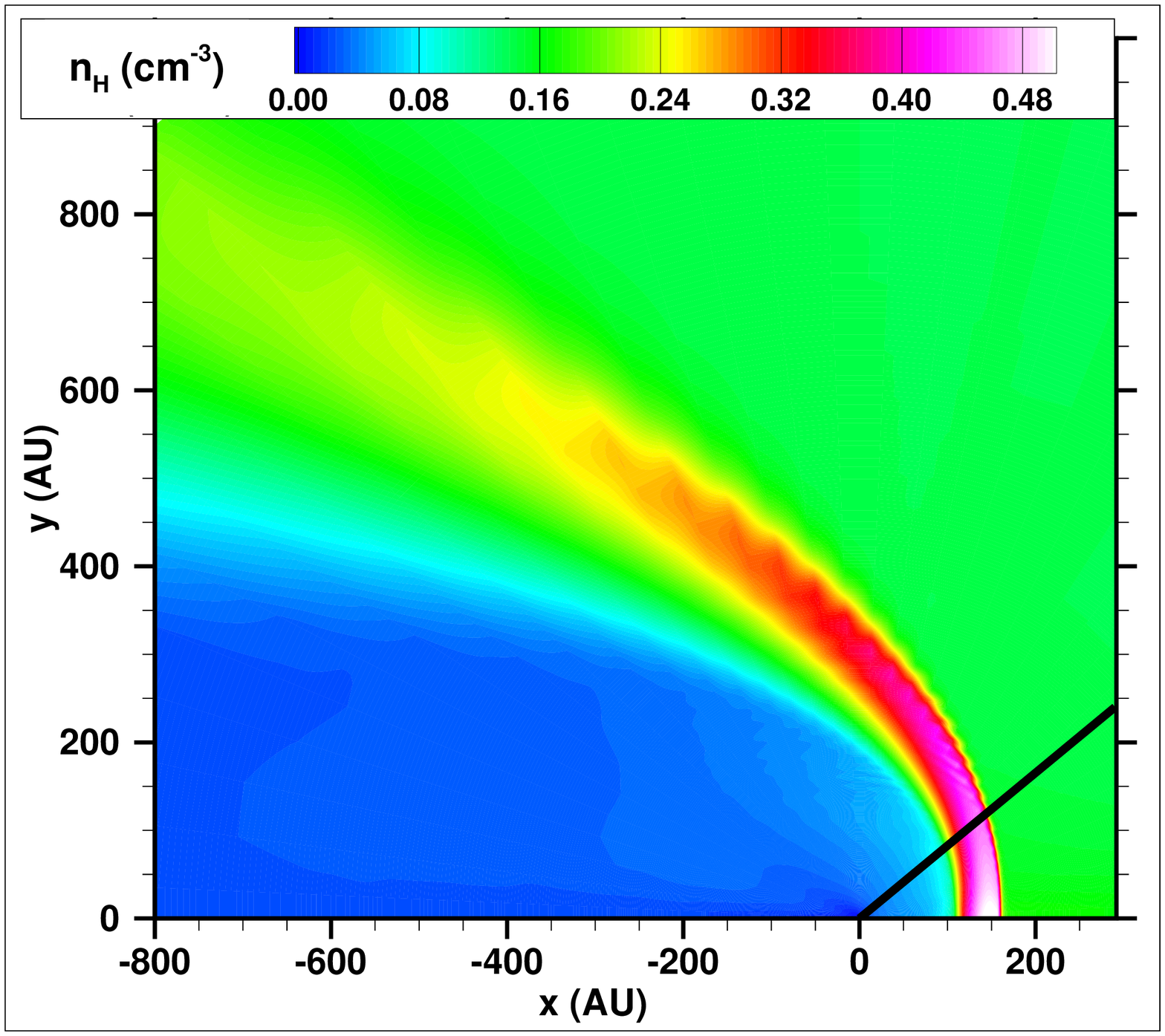}{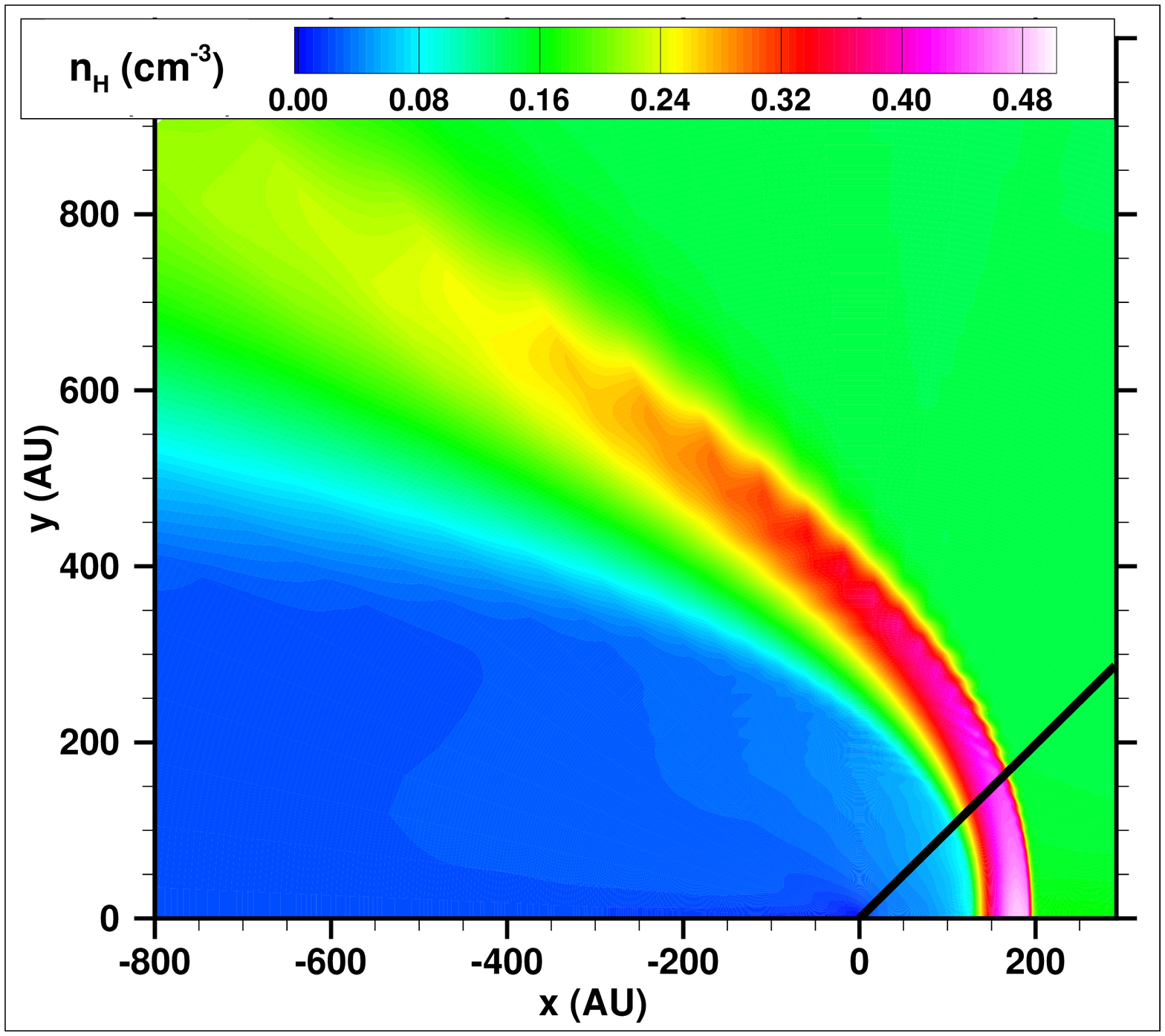}
\label{fig9}
\caption{Neutral hydrogen number density distributions surrounding HD~9826 ({\it left}) and HD192310 ({\it right}), each located at $x=y=0$ AU. The LISM flow is coming in from the right. {\em Left:} density of hot \ion{H}{1} 
surrounding HD~9826, with the line of sight from the Sun to HD~9826 at
39.6$^{\circ}$ indicated by the black line. The hydrogen wall is
clearly seen from the enhancement of the hydrogen density near and trailing from the leading nose of the astrosphere. {\em Right:} Same for 
HD~192310, with the line of sight from the Sun to HD~192310 at an angle of 44.7$^{\circ}$.}
\end{figure}

\begin{figure}
\figurenum{10}
\epsscale{0.9}
\plotone{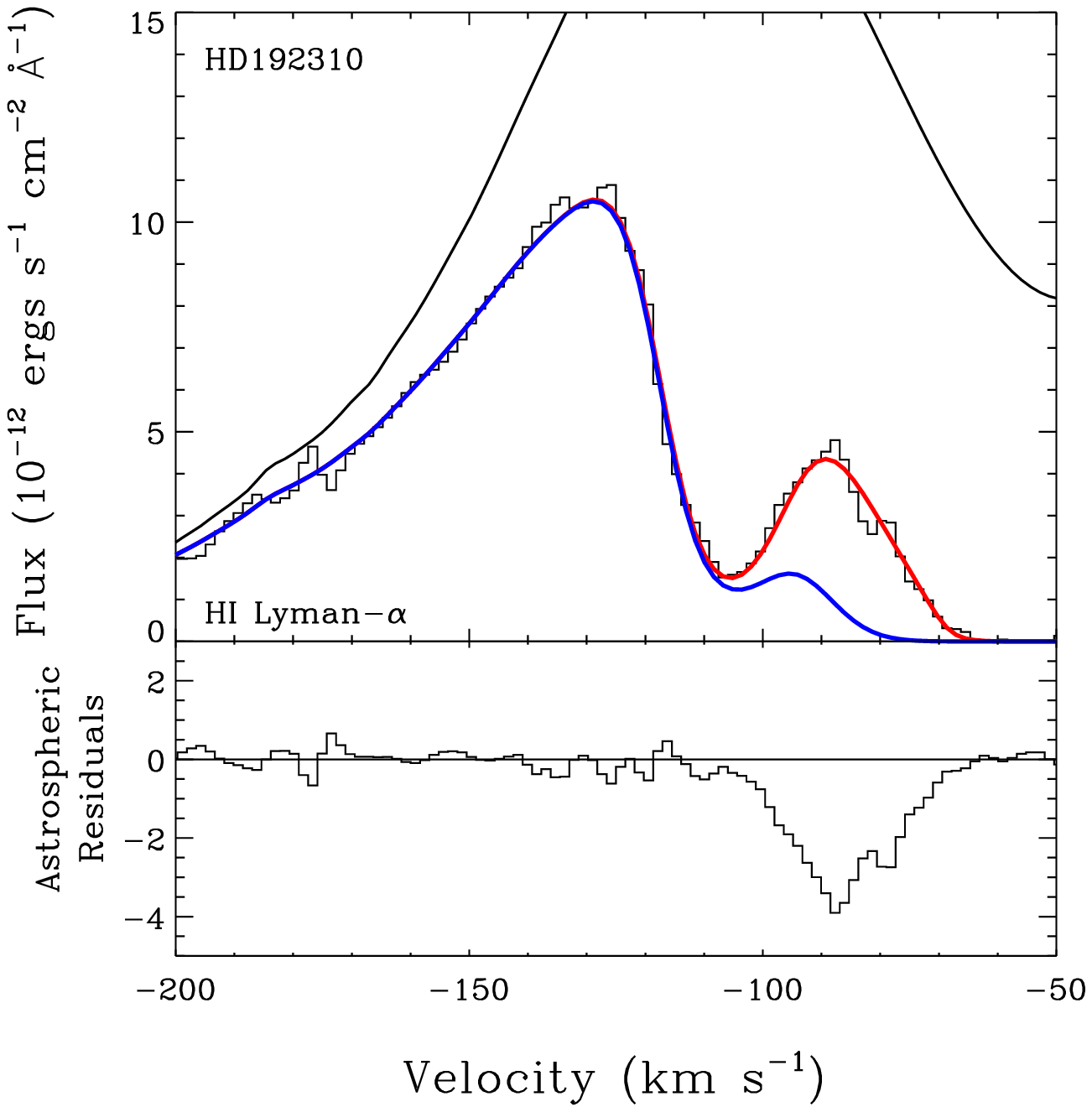}
\label{fig10}
\caption{Expanded version of Figure~\ref{fig5} showing the short wavelength
side of the reconstructed stellar Lyman-$\alpha$ emission (solid black
line) and the observed spectrum (histogram). The red line shows the
predicted Lyman-$\alpha$ stellar emission with absorption by the three
interstellar components. The blue line shows the effect of including
additional absorption from the stellar astrosphere  
calculated assuming a mass-loss rate of
5.9 $\dot{M}_\odot$. 
Inclusion of this amount of astrosphere absorption
produces very large residuals centered at --90 km~s$^{-1}$.}
\end{figure}

\begin{figure}
\figurenum{11}
\epsscale{0.9}
\plotone{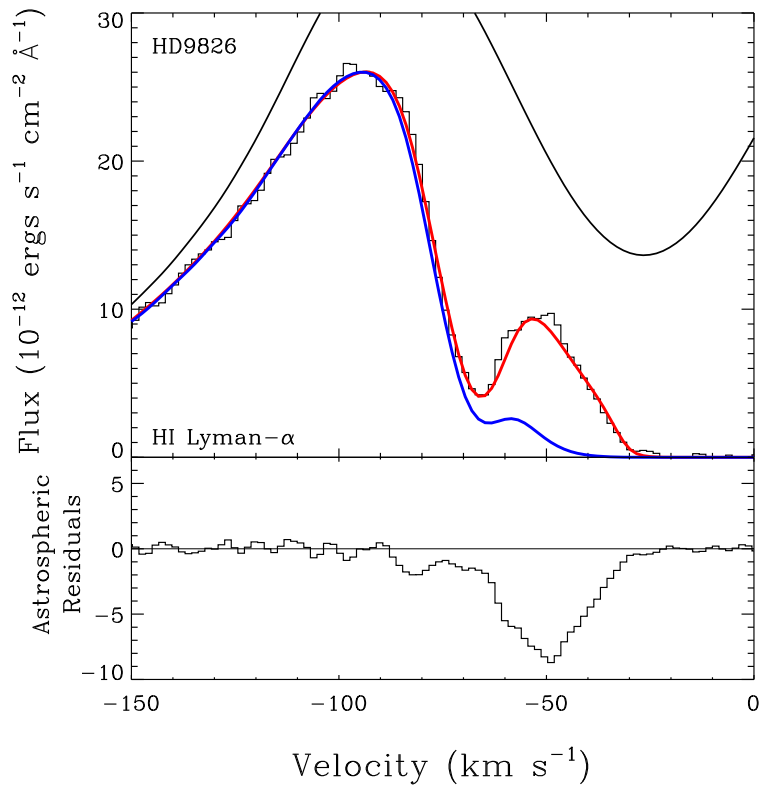}
\label{fig11}
\caption{Same as Figure~\ref{fig10} but for HD~9826 with a mass-loss rate of 5.1 $\dot{M}_\odot$. The residuals associated
  with the inclusion of astrosphere absorption are centered at --50 km s$^{-1}$.}
\end{figure}

Another test for astrospheric absorption is to compare the relative velocities
of interstellar \ion{H}{1} and \ion{D}{1} absorption. Note that the rest 
wavelengths for the \ion{H}{1} and \ion{D}{1} lines automatically include the 
theoretical wavelength difference 
of 0.3307~\AA, corresponding to 81.6 km~s$^{-1}$. If a star has a
hydrogen wall, then the central velocity of \ion{H}{1} absorption (produced by
interstellar and astrospheric absorption) will be shifted
to shorter wavelengths (blueshifted). The velocity difference, $\Delta v = v($\ion{H}{1}$) - v($\ion{D}{1}$)$, is typically $-3$\,km\,s$^{-1}$ \citep[e.g.,][]{wood96}. On the other hand, if there is no
astrospheric absorption and the line of sight traverses
the hydrogen wall around the Sun, then the velocity difference will be
positive (redshifted) due to the heliospheric absorption, resulting in a $\Delta v \approx 3$\,km\,s$^{-1}$. 
We find that the observed velocity difference average ($\Delta v$) for the three
interstellar components is 2.93 km~s$^{-1}$ for
HD~192310, +0.78 km~s$^{-1}$ for HD~9826, and --1.79 km~s$^{-1}$ for
HD~206860. These small velocity differences could be explained by
systematic errors in the fitting procedure resulting from the need to simultaneously 
analyze the three blended components along the line of sight resulting in 12 absorbing columns for each Lyman-$\alpha$ line
profile. However, the velocity difference for HD~192310 does approach 3\,km\,s$^{-1}$, and therefore indicates possible heliospheric absorption. \citet{Wood2005b} provides evidence that HD 192310's total $\log N$(\ion{H}{1}) value may be low enough for detectable heliosphere absorption, which could warrant further observation and analysis of this sight line. 

There are several possible explanations for the absence of a
measurable astrospheric absorption signature in the Lyman-$\alpha$
line profiles. If the interstellar hydrogen column density $\log N$(\ion{H}{1})
$> 18.7$, then the very broad Lyman-$\alpha$ absorption would 
completely absorb the underlying astrospheric absorption.
This is not the case for these three sight lines since the sum of the 
three absorption components does not exceed
$\log N$(\ion{H}{1}) = 18.4 for any sight line. 

A second possibility for nondetection could be 
that the line of sight to the star is at a large angle with respect to
the nose of the incoming interstellar flow. The
maximum velocity difference occurs when observing at the same angle as
the inflow. At a tangential angle the astrospheric velocity difference
is close to zero, and when observing close to the tail (180$^{\circ}$
from the inflow direction) the absorption by the hydrogen gas flowing
past the star is likely redshifted relative to the interstellar velocity 
\citep{Wood2014a}. Figure~\ref{fig9} shows plasma models for the astrospheres
of HD~9826 and HD~192310, showing the location of the
hydrogen walls. These simulations are multifluid, axisymmetric, hydrodynamic models treating the charge exchange of interstellar neutrals with protons in the stellar wind, originally developed for the heliosphere. The models assume that hydrogen is 
42\% ionized, similar to what has been determined for the LISM near the 
Sun. Our multifluid treatment successfully approximates the non-Maxwellian distribution of neutral hydrogen associated with the interaction of the heliosphere with the surrounding local interstellar medium. Further details regarding this type of model can be found in \citet{zank96}, \citet{zank99}, \citet{Heer2008}, \citet{mueller06}, and \citet{pogorelov2017}. The observed sight lines (indicated in the figure) are
39.6$^{\circ}$ for HD~9826 and 44.7$^{\circ}$ for HD~192310,
respectively, relative to the inflow nose. Table~\ref{astroprop} provides a more detailed look at the model parameters, and Figures~\ref{fig10} and \ref{fig11} show the expected astrospheric absorption based on these models for HD~192310 and HD~9826, respectively, assuming stellar mass loss rates of 5.9 $\dot{M}_\odot$ for HD~192310 and 5.1 $\dot{M}_\odot$ for HD~9826. These mass loss rates are based on the stellar X-ray luminosities and the correlation with mass loss identified by \citet{Wood2005b} (with a recent version in \citet{Wood2014b}). If hydrogen walls are
present, they should have been detected. 

\begin{deluxetable}{lccccccc}
\tablewidth{0pt}
\tablecolumns{8}
\tablecaption{Model Astrosphere\tablenotemark{a} Parameters\label{astroprop}}
\tablehead{
\colhead{Star}&\colhead{$v$(ISM)}&\colhead{Ionization}&\colhead{$n$(ISM)}&\colhead{Sightline}&\colhead{$v$(Stellar Wind)}&\colhead{$n$(Stellar Wind)}&\colhead{Stellar Mass}\\
\colhead{Name}&\colhead{}&\colhead{Fraction (ISM)}&\colhead{Total}&\colhead{Angle}&\colhead{at 1 AU}&\colhead{at 1 AU}&\colhead{Loss Rate}\\
\colhead{}&\colhead{(km s$^{-1}$)}&\colhead{}&\colhead{(cm$^{-3}$)}&\colhead{($\theta$)}&\colhead{(km s$^{-1}$)}&\colhead{(cm$^{-3}$)}&\colhead{($\dot{M}_\odot$)}\\
}
\startdata
HD~9826& 56.0 & 42\% & 0.24 & 39.9 & 400 & 25.3 & 5.1\\
HD~192310 & 52.0 &42\% & 0.24 & 44.7 & 400 & 29.4 & 5.9
\enddata
\tablenotetext{a}{These are axisymmetric, multifluid, hydrodynamical models that assume a temperature of 100,000 $K$ at 1 AU for the stellar wind, and 8,000 $K$ for the ISM. A solar wind density of 5 cm$^{-3}$ and velocity of 400 km s$^{-1}$ at 1 AU were used as a basis for expressing the mass loss rates of these solar-like stars.}
\end{deluxetable}

The third possibility is that the stars are not located inside of
interstellar clouds containing neutral hydrogen gas. The vast majority of analyzed stars of distances greater than 10 pc yield non-detections of astrospheric absorption \citep{Wood2005b}, which is likely due to stars being surrounded by ionized plasma within the Local Bubble. In this case, the hydrogen wall structure will be significantly different, if present at all \citep{mueller06}, and thus
no astrospheric signatures are expected in the Lyman-$\alpha$ line profiles. We
believe that this is the most likely explanation for the
non-detections because the low column density and favorable observing angles are both conducive for detection.

\section{Conclusions}
\label{conclusion}

Our analysis of the high-resolution STIS spectra of three nearby
planet-hosting stars allowed us to identify and measure the properties of nine interstellar
absorption components along their lines of sight. We can reliably assign eight of the nine components
with interstellar clouds previously identified by \citet{Redfield2008}
on the basis of location inside or very near the cloud boundary and
radial velocity within 2.0~km~s$^{-1}$ of that predicted from the
cloud's velocity vector. In all eight cases, the measured and predicted
temperatures agree within the admittedly large errors and the turbulence
and depletions agree in most cases. 

None of the three stars show any direct evidence of astrosphere absorption,
implying that these stars are likely not located inside of the partially
ionized clouds observed along their lines of sight. Instead, these
stars are likely surrounded by fully ionized interstellar gas that produces very different or
no hydrogen walls. We cannot, therefore, estimate the mass-loss 
rates for these three planet-hosting stars by the astrosphere technique.

We also predict the interstellar radial velocities of clouds located along 
the lines of sight to all known planet-hosting stars within 20~pc of the Sun. 
Although our knowledge of the clouds and their kinematic properties is incomplete, the predicted velocities should prove useful
in analyzing low-resolution spectra of these stars to search for weak
absorption or emission from their exoplanets. Also, analysis of
stellar winds using the astrosphere technique requires knowledge of the
interstellar flow vector. 

Alongside the kinematics, measuring the densities of various ions within the clouds is a necessary step when interpreting spectroscopic measurements commonly used in exoplanet characterization. For example, without accounting for Lyman-$\alpha$ absorption within the ISM along the sight line, it is impossible to accurately characterize exoplanet mass loss during and after transits by analyzing line of sight spectra. 

Without careful analysis of the many features of the local interstellar medium, our picture of our local environment remains incomplete. Therefore, as our characterization and scrutiny of exoplanetary atmospheres becomes more precise over the coming decades, it will be critical to have detailed observations and knowledge of the intervening and surrounding interstellar medium.

\acknowledgments
We acknowledge support by {\it HST} Grants GO-12475 and GO-12596 awarded by the Space 
Telescope Science Institute, which is operated by the Association 
of Universities for Research in Astronomy, Inc., for NASA, under 
contract NAS 5-26555. EE gives thanks for the student fellowship 
awarded from the Connecticut Space Grant Consortium in support of 
this research.

This work has made use of data from the European Space Agency (ESA) mission
{\it Gaia} (\url{https://www.cosmos.esa.int/gaia}), processed by the {\it Gaia}
Data Processing and Analysis Consortium (DPAC,
\url{https://www.cosmos.esa.int/web/gaia/dpac/consortium}). Funding for the DPAC
has been provided by national institutions, in particular the institutions
participating in the {\it Gaia} Multilateral Agreement.

\facility{{\it HST} (STIS)}

\startlongtable
\begin{deluxetable}{lccccccccccccc}
\tabletypesize{\tiny}

\tablecaption{LISM Properties for the Sightlines to Planet-Hosting
  Stars within 20 pc of the Sun\tablenotemark{a}\label{planettable}}
\tablewidth{0pt}
\tablecolumns{1}
\tablehead{
\multicolumn{2}{c}{Star} & \colhead{$d$\tablenotemark{b}} & \colhead{$l$} & \colhead{$b$} &
\colhead{m$_v$} & \colhead{LISM\tablenotemark{c}} &
\multicolumn{2}{c}{\underline{Component 1}} & 
\multicolumn{2}{c}{\underline{Component 2}} &
\multicolumn{2}{c}{\underline{Component 3}} & Ref\tablenotemark{d}\\
\colhead{} & \colhead{} & \colhead{(pc)} & \colhead{($^{\circ}$)} &
\colhead{($^{\circ}$)} & \colhead{} & \colhead{clouds} &
\colhead{$v_{\rm rad}$} & \colhead{$N(\rm H I)$}
&  \colhead{$v_{\rm rad}$} & \colhead{$N(\rm H I)$} &  
\colhead{$v_{\rm rad}$} & \colhead{$N(\rm H I)$} & \colhead{}}
\startdata
Proxima Cen & GJ~551 & 1.30 & 313.94 & --1.93 & 11.13 & G & --18.1 & 
17.6 & & & & & 3\\
Lalande 21185 & GJ~411 & 2.55\tablenotemark{e} & 185.12 & +65.43 & 7.52 & LIC & (4.66) & (17.10)
 & & & & & 8 \\
$\epsilon$ Eri & GJ~144 & 3.20 & 195.85 & --48.05 & 3.73 & LIC & 16.1 & 
17.93 & & & & & 6\\
Ross 128 & GJ~447 & 3.37 & 270.15 & +59.56 & 11.15 & Leo,NGP & (--0.86) & & (0.13) & & & & 8 \\
HD~1326 & GJ~15A & 3.56 & 116.68 & --18.45 & 8.13 & LIC,Hyades &
(9.20) & (18.06) & (10.06) & & & & 8\\
$\tau$~Cet & GJ~71 & 3.60 & 173.10 & --73.44 & 3.50 & LIC & 12.34 &
18.0 & & & & & 2\\ 
YZ~Cet & GJ~54.1 & 3.71 & 149.71 & --78.76 & 12.07 & LIC,Mic,Cet &
(9.00) & (17.99) & (4.92) & & (16.75) & & 8\\
HIP~36208 & GJ~273 & 3.80\tablenotemark{f} & 212.34 & +10.37 & 9.87 & LIC,Aur,Gem &
(19.65) & (17.86) & (22.52) & & (35.42) & & 8\\
Kapteyn & GJ~191 & 3.93 & 250.53 & --36.00 & 8.85 & Blue,Dor & (10.40)
& & (21.48) & & & & 8\\ 
Wolf~1061 & GJ~628 & 4.31 & 3.35 & +23.68 & 10.07 & G,Mic & (--29.55) &
& (--25.15) & & & & 8\\
HIP~86162 & GJ~687 & 4.55 & 98.60 & +31.96 & 9.15 & LIC & (--2.32) &
(17.39) & & & & & 8\\ 
HIP~85523 & GJ~674 & 4.55 & 343.00 & --6.78 & 9.41 & G,Aql & (--24.37) & &
(--28.24) & & & & 8\\ 
IL~Aqr & GJ~876 & 4.68 & 52.00 & --59.63 & 10.19 & LIC & 2.8 & 18.03 & 
& & & & 6\\
HD~204961 & GJ~832 & 4.97 & 349.17 & --46.35 & 8.67 & LIC & --17.1 & 
18.20 & & & & & 6\\
HD~26965 & GJ~166A & 5.04 & 200.75 & --38.05 & 4.43 & LIC & 21.73 & 17.8
 & & & & & 2 \\
 & GJ~3323 & 5.38 & 206.43 & --27.47 & 12.20 & LIC,Aur,Hyades &
 (21.97) & (17.89) & (24.65) & & (11.69) & & 8\\
HD~20794 & GJ~139 & 6.00 & 250.75 & --56.08 & 4.27 & Dor,Blue,G &
(31.42) & & (9.33) & & (14.87) & & 8\\ 
HIP 74995 & GJ~581 & 6.30 & 354.08 & +40.02 & 10.56 & G,Gem &
--24.1\tablenotemark{g} & 18.01\tablenotemark{g} & (--27.57) & & & & 2,8\\
HIP~80459 & GJ~625 & 6.47 & 83.21 & +42.79 & 10.17 & LIC,Oph,NGP &
(--7.77) & (16.87) & (--16.28) & & (--10.08) & & 8\\
HD~219134\tablenotemark{h} & GJ~892 & 6.53 & 109.90 & --3.20 & 5.57 & LIC & (5.53) &
(18.03) & & & & & 8\\ 
HD~156384 & GJ~667C & 7.25 & 351.84 & +1.42 & 5.89 & G & --22.5 & 
17.98 & & & & & 6\\ 
Fomalhaut & GJ~881 & 7.70\tablenotemark{f} & 20.49 & --64.91 & 1.16 & LIC,Mic & --5.87
& (17.40) & --10.64 & & & & 2,8\\
HIP~86287 & GJ~686 & 8.16 & 42.24 & +24.30 & 9.58 & LIC,Mic,Oph & (--19.54) & (16.39) 
& (-25.11) & & (-29.11) & & 8\\ 
61 Vir & GJ~506 & 8.51 & 311.86 & +44.09 & 4.74 & NGP,Leo & --14.74 & 17.9
& (--11.01) & & & & 2,7\\ 
HD~192310 & GJ 785 & 8.80 & 15.62 & --29.40 & 5.72 & Vel,Mic,LIC & --30.40 & 
17.96 & --24.24 & 17.90 & --18.18 & 16.17 & 1\\
HIP~109388 & GJ~849 & 8.80 & 55.89 & --45.40 & 10.37 & LIC & (--6.80)
& (17.56) & & & & & 8\\ 
HIP~56528 & GJ~433 & 9.07 & 284.88 & +27.65 & 9.81 & Gem & (6.51) & &
& & & & 8\\ 
HD~102365 & GJ~442A & 9.29 & 289.80 & +20.71 & 4.88 & G & (--10.52) &
& & & & & 8\\ 
HD~285968 & GJ~176 & 9.47 & 180.02 & --17.43 & 9.51 & LIC,Hyades,Aur & 
29.0\tablenotemark{g} & 17.46\tablenotemark{g} & (12.9) & & (21.71) & & 6,8\\
Ross~905\tablenotemark{h} & GJ~436 & 9.76 & 210.54 & +74.56 & 10.61 & ? & --4.1 & 18.04 & 
& & & & 6\\
WISE~J1217+1626 & & 10.10\tablenotemark{i} & 265.24 & +76.80 & & NGP,Leo &
(--1.28) & & (--2.07) & & & & 8\\
LHS~3776 & GJ~1265 & 10.26 & 39.08 & --52.51 & 13.67 & LIC & (-7.61)
(17.35) & & & & & & 8\\
$\beta$~Gem & GJ~286 & 10.36\tablenotemark{f} & 192.23 & +23.41 & 1.14 & LIC,Gem & 19.65
& 18.0 & 31.84 & 17.8 & & & 2\\ 
HIP~83043 & GJ~649 & 10.38 & 46.52 & +35.34 & 9.66 & NGP,Oph,LIC &
(--26.10) & & (--25.73) & & (--17.78) & (16.42) & 8\\ 
HD~122303 & GJ~536 & 10.41 & 335.06 & +55.81 & 9.71 & NGP,Leo,Gem &
(--19.89) & & (--13.53) & & (--13.15) & & 8\\
HD~147379 & GJ~617A & 10.77 & 99.90 & +39.47 & 8.90 & LIC & (--2.55)
 & (17.28) &  & &  & & 8\\
HD~13445 & GJ~86 & 10.79 & 275.93 & --61.96 & 6.17 & Cet,Dor,Vel &
(13.44) & & (26.73) & & (16.78) & & 8\\
HD~57050 & GJ~1148 & 11.02 & 210.34 & +9.26 & 8.71 & LIC,Aur & (20.16) & (17.89)
& (22.72) & & & & 8\\ 
HD~3651 & GJ~27 & 11.14 & 119.17 & --41.53 & 5.88 & LIC,Hyades &
(10.24) & & (12.32) & & & & 8\\ 
HD~85512 & GJ~370 & 11.28 & 271.68 & +8.16 & 7.65 & G,Cet & --3.3\tablenotemark{g} 
& 18.43\tablenotemark{g} & (14.3) & & & & 6,8\\
HIP~63510 & GJ~494 & 11.51 & 311.83 & +75.09 & 9.75 & NGP & (-8.39) & &
& & & & 8\\
HIP~11048 & GJ~96 & 11.94 & 138.29 & --12.27 & 9.35 & LIC & (16.18) & (18.27) &
& & & & 8\\
HIP~22627 & GJ~179 & 12.36 & 192.23 & --22.86 & 12.02 & LIC,Aur,Hya &
(23.46) & (18.01) & (23.75) & & (12.64) & & 8\\ 
Trappist-1\tablenotemark{h} & & 12.43 & 69.71 & --56.64 & 18.80 & LIC,Cet & (--1.25) &
(17.86) & (--12.21) & & & & 8\\ 
HD~69830 & GJ~302 & 12.56 & 234.56 & +12.82 & 5.95 & LIC,Aur & (14.04)
& (17.43) & (20.21) & & & & 8\\
55 Cnc\tablenotemark{h} & GJ~324A & 12.59 & 196.79 & +37.70 & 5.95 & LIC,Gem & (14.75) &
(17.78) & (27.78) & & & & 8\\
 & GJ~1132\tablenotemark{h} & 12.62 & 277.26 & +7.76 & 13.52 & G,Cet & (-2.73) & &
(8.82) & & & & 8\\
VHS~J1256-1257 & & 12.7\tablenotemark{j} & 304.67 & +49.90 & 17.76 & 
Leo & (-8.71) & & & & && 8\\
HD~147513 & GJ~9559 & 12.91 & 341.62 & +7.21 & 5.38 & G,Gem &
(--26.63) & & (--25.30) & & & & 8\\ 
HD~40307 & GJ~2046 & 12.94 & 268.81 & --30.34 & 7.15 & Blue,G,Vel & 9.4 & 
18.60 & (7.63) & & (18.38) & & 6,8\\
HD~9826 & GJ 61 & 13.41 & 132.00 & --20.67 & 10.07 & ?,Hyades,LIC
& 9.11 & 17.78 & 12.14 & 17.55 & 16.45 & 17.34 & 1\\
$\gamma$~Cep & GJ~903 & 13.54 & 118.99 & +15.32 & 3.22 & LIC & (8.33) &
(18.04) & & & & & 8\\ 
Ross 1020 & GJ~3779 & 13.75 & 14.12 & +82.43 & 13.01 & NGP & (-8.29) &
& & & & & 8\\ 
47 UMa & GJ~407 & 13.80 & 175.78 & +63.37 & 5.04 & LIC,NGP & (5.31) &
(17.23) & (12.91) & & & & 8\\ 
& GJ~1214\tablenotemark{h} & 14.65 & 26.16 & +23.61 & 14.67 & Mic,G,Oph & --26.4\tablenotemark{g} & 
18.06\tablenotemark{g} & (--27.76) & & (--28.78) & & 6,8\\
HIP~79431 & & 14.54 & 355.20 & +22.97 & 11.37 & G,Gem & (--29.24) & &
(--28.63) & & & & 8\\ 
HD~136352 & GJ~582 & 14.69 & 327.08 & --7.38 & 5.65 & G & (--20.49) & &
& & & & 8\\ 
LHS~3844 &  & 14.89 & 318.25 & --43.91 &  & Vel & (-7.31) & &
& & & & 8\\ 
HIP~49189 & GJ~378 & 14.96 & 168.73 & +51.19 & 10.14 & LIC & (9.56) & (17.63) &
& & & & 8\\ 
LHS~1140\tablenotemark{h} & GJ~3053 & 14.99 & 115.40 & --78.05 & 14.15 & LIC,Mic,Cet &
(6.90) & (18.01) & (1.85) & & (10.67) & & 8\\ 
HIP~19394 & GJ~163 & 15.14 & 262.78 & --45.32 & 11.8 & Dor,G,Blue &
(21.29) & & (11.36) & & (8.56) & & 8\\ 
LHS~2037 & GJ~317 & 15.20 & 246.80 & +11.19 & 11.98 & G,Cet & (10.61) &
& (35.87) & & & & 8\\ 
HD~38858 & GJ~1085 & 15.25 & 209.38 & --15.84 & 5.97 & LIC,Aur &
(22.16) & (17.90) & (25.19) & & & & 8\\ 
51 Peg & GJ~882 & 15.47 & 90.06 & --34.73 & 5.46 & Eri,Hyades &
(--1.23) & & (8.11) & & & & 8\\ 
HD~160691 & GJ~691 & 15.61 & 340.06 & --11.50 & 5.15 & G,Aql &
(--22.64) & & (--23.30) & & & & 8\\
$\tau$~Boo & GJ~527 & 15.66 & 358.94 & +73.89 & 4.49 & NGP,Gem &
--11.64 & & (--9.60) & & & & 7,8\\
HD~190360 & GJ~9683 & 16.01 & 67.41 & --0.67 & 5.71 & Mic,G,Aql &
(--20.27) & & (--12.49) & & (--17.76) & & 8\\ 
HIP~85647 & GJ~676A & 16.03 & 339.10 & --9.54 & 9.59 & G,Aql & (--22.96) & 
& (--24.75) & & & & 8\\ 
HD~113538 & GJ~9425 & 16.29 & 305.03 & +10.37 & 9.06 & G & (--15.72) & 
& & & & & 8\\ 
HD~21749 & GJ~143 & 16.33 & 279.23 & --45.82 & 8.14 & G,Cet,Vel & (5.88) & 
& (12.19) & & (13.98) & & 8\\ 
HD~128311 & GJ~3860 & 16.34 & 2.73 & +59.84 & 7.45 & Gem,Leo,NGP &
(--17.25) & & (--14.68) & & (--21.38) & & 8\\
HD~177565 & GJ~744 & 16.93 & 359.45 & --19.08 & 6.16 & Aql,Vel,Mic &
(--19.85) & & (--36.07) & & (--24.07) & & 8\\
HIP~79126 & GJ~3942 & 16.94 & 82.20 & +45.47 & 10.25 & NGP & (--10.29) & 
& & & & & 8\\ 
HD~7924 & GJ~9054 & 17.00 & 124.74 & +13.95 & 7.19 & LIC & (9.21) &
(18.14) & & & & & 8\\ 
HD~17051 & GJ~108 & 17.33 & 268.81 & --58.32 & 5.40 & G,Dor,Vel &
(10.31) & & (26.61) & & (19.53) & & 8\\
HD~10647 & GJ~3109 & 17.34 & 286.90 & --61.77 & 5.52 & Dor,Vel,Cet &
(23.93) & & (12.70) & & (8.12) & & 8\\ 
$\rho$~Cor~Bor & GJ~9537 & 17.48 & 53.49 & +48.92 & 5.41 & NGP,Mic,Oph
& (--20.13) & & (--17.32) & & (--20.05) & & 8\\
HD~1237A & GJ~3021 & 17.56 & 304.87 & --37.14 & 6.58 & G,Vel,Cet &
(--4.88) & & (--3.14) & & (--8.93) & & 8\\
$\epsilon$~Vir & GJ~504 & 17.54 & 322.79 & +71.31 & 5.22 & NGP,Leo &
(--11.35) & & (--8.37) & & & & 8\\ 
HD~189567 & GJ~776 & 17.91 & 328.47 & --31.99 & 6.07 & G,Vel &
(--13.49) & & (--18.32) & & & & 8\\
70 Vir & GJ~9446 & 17.91 & 337.67 & +74.10 & 4.97 & NGP,Leo & (--11.89)
& & (--8.79) & & & & 8\\ 
14 Her & GJ~614 & 17.94 & 69.17 & +46.94 & 6.67 & NGP,Oph & (--15.36) &
& (--18.45) & & & & 8\\ 
HD~206860 & GJ 9751 & 18.13 & 69.86 & --28.27 & 5.95 & Vel,Eri,LIC & --14.68 & 
17.35 & --8.00 & 17.79 & --5.44 & 18.14 & 1\\
HIP~84460 & GJ~3998 & 18.16 & 32.27 & +26.15 & 10.83 & LIC,G,Mic &
(--21.27) & (16.30) & (26.60) & & (--25.88) & & 8\\
83 Leo & GJ~429 & 18.21 & 259.28 & +58.51 & 7.53 & Leo,NGP & (1.45) & &
(3.78) & & & & 8\\ 
HD~99492 & GJ~429B & 18.21 & 259.28 & +58.51 & 7.53 & Leo,NGP &
(1.45) & & (3.78) & & & & 8\\
HD~154088 & GJ~652 & 18.28 & 355.24 & +7.67 & 6.58 & G & (--28.49) & &
& & & & 8\\ 
HD~27442 & GJ~9153 & 18.28 & 270.20 & --42.56 & 4.44 & G,Vel,Cet &
(8.57) & & (18.67) & & (18.86) & & 8\\ 
HD~39091 & GJ~9189 & 18.28 & 292.51 & --29.78 & 5.67 & G,Vel,Cet &
(--2.26) & & (2.28) & & (-0.33) & & 8\\
HD~154345 & GJ~651 & 18.29 & 73.02 & +37.66 & 6.74 & Oph,NGP,LIC &
(--20.55) & & (--15.27) & & (--10.80) & (16.73) & 8\\ 
HD~87883 & & 18.30 & 191.19 & +54.54 & 7.55 & LIC,Leo & (8.97) &
(17.46) & (10.51) & & & & 8\\ 
WD~0806-661 & GJ~3483 & 19.26 & 279.42 & --17.57 & 13.70 & G,Blue,Cet &
(0.88) & & (4.96) & & (10.31) & & 8\\ 
HD~192263 & & 19.65 & 41.87 & --18.70 & 7.77 & Eri,Aql,Mic & (--21.58)
& & (--36.25) & & (--26.85) & & 8\\ 
$\beta$~Pic & GJ~219 & 19.75 & 258.36 & --30.61 & 3.86 & Blue,G &
(9.32) & & (11.93) & & & & 8\\ 
HD~189733\tablenotemark{h} & GJ~4130 & 19.78 & 60.96 & --3.92 & 7.65 & Aql,Eri, Mic &
(--18.64) & & (--15.90) & & (--22.23) & & 8\\
7 CMa & GJ~9214 & 19.82 & 228.69 & --11.80 & 3.91 & LIC,Blue & (18.08)
& (17.68) & (12.66) & & & & 8\\
\enddata
\tablenotetext{a}{Data in parentheses are predicted from models. The velocities from \citet{Redfield2008}: \url{http://lism.wesleyan.edu/LISMdynamics.html} and column densities from \citet{Redfield2000}: \url{http://lism.wesleyan.edu/ColoradoLIC.html}.}
\tablenotetext{b}{\textbf{Distances are from GAIA DR2 \citep{Gaia2016, Gaia2018} unless otherwise noted.}}
\tablenotetext{c}{Interstellar partially ionized clouds in the line of
  sight to the star. For lines of sight with predicted interstellar
  velocities, the first listed cloud is predicted to be the dominant absorber.}
\tablenotetext{d}{(1) This paper, (2) \citet{Redfield2008}, 
(3) \citet{Wood2001}, (4) \citet{Wood2014a}, (5) \citet{Wood2014b},
(6) \citet{Youngblood2016}, (7) \citet{Malamut2014}, (8) velocities from \url{http://lism.wesleyan.edu/LISMdynamics.html} and column densities from \url{http://lism.wesleyan.edu/ColoradoLIC.html}.}
\tablenotetext{e}{\citet{van2009}}
\tablenotetext{f}{\citet{Hipparcos}}
\tablenotetext{g}{N(H~I), actually log N(H~I), is the sum of all
  velocity components, and
  $v_{\rm rad}$ is the measured mean velocity when several velocity
  components are likely present in this line of sight.}
  \tablenotetext{h}{Hosts a transiting exoplanet.}
  \tablenotetext{i}{\citet{Dupuy2013}}
  \tablenotetext{j}{\citet{gauza2015}}

\end{deluxetable}

\end{document}